\documentclass[10pt,preprint]{aastex}
\newcommand{\hmpcC}{h^{-3}{\rm\,Mpc^3}}
\newcommand{\hgpcC}{h^{-3}{\rm\,Gpc^3}}

\newcommand{\kmsmpc}{{\rm\ km\ s^{-1}\ Mpc^{-1}}}
\newcommand{\kms}{\ensuremath{\rm\ km\ s^{-1}}}

\newcommand{\photo}{{\tt photo}}
\newcommand{\target}{{\tt target}}
\newcommand{\spectro}{{\tt spectro}}
\newcommand{\us}{\ensuremath{u^*}}
\newcommand{\gs}{\ensuremath{g^*}}
\newcommand{\rs}{\ensuremath{r^*}}
\newcommand{\is}{\ensuremath{i^*}}
\newcommand{\zs}{\ensuremath{z^*}}
\newcommand{\rpet}{\ensuremath{r^*_{\rm Petro}}}
\newcommand{\gpet}{\ensuremath{g^*_{\rm Petro}}}
\newcommand{\rpsf}{\ensuremath{r^*_{\rm psf}}}
\newcommand{\rmodel}{\ensuremath{r^*_{\rm model}}}
\newcommand{\sbr}{\ensuremath{\mu_{r^*,\rm Petro}}}
\newcommand{\cperp}{\ensuremath{c_\perp}}
\newcommand{\cpar}{\ensuremath{c_\parallel}}
\newcommand{\cparcoeff}{\ensuremath{C}}

\newcommand{\tableskip}{\\[-6pt]}
\newlength{\tskip}
\newlength{\colwidth}
\newcommand{\colskip}{@{\hspace{0.5in}}}

\newcommand{\beq}{\begin{equation}}
\newcommand{\eeq}{\end{equation}}
\newcommand{\beqa}{\begin{eqnarray}}
\newcommand{\eeqa}{\end{eqnarray}}

\epsscale{0.6}    

\begin{document}
\title{Spectroscopic Target Selection for the Sloan Digital Sky Survey:\\
The Luminous Red Galaxy Sample}
\author{%
Daniel~J.~Eisenstein\altaffilmark{1,2,3,19},
James~Annis\altaffilmark{4}, 
James~E.~Gunn\altaffilmark{5},
Alexander~S.~Szalay\altaffilmark{6},
Andrew~J.~Connolly\altaffilmark{7},
R.~C.~Nichol\altaffilmark{8},
Neta~A.~Bahcall\altaffilmark{5},
Mariangela~Bernardi\altaffilmark{2},
Scott~Burles\altaffilmark{4},
Francisco~J.~Castander\altaffilmark{9,10},
Masataka~Fukugita\altaffilmark{1,11},
David~W.~Hogg\altaffilmark{1,12,19},
\v{Z}eljko~Ivezi\'c\altaffilmark{5},
G.~R.~Knapp\altaffilmark{5},
Robert~H.~Lupton\altaffilmark{5},
Vijay~Narayanan\altaffilmark{5},
Marc~Postman\altaffilmark{13},
Daniel~E.~Reichart\altaffilmark{2,14,19},
Michael~Richmond\altaffilmark{15}
Donald~P.~Schneider\altaffilmark{16},
David~J.~Schlegel\altaffilmark{5},
Michael~A.~Strauss\altaffilmark{5},
Mark~SubbaRao\altaffilmark{2},
Douglas~L.~Tucker\altaffilmark{5},
Daniel~Vanden~Berk\altaffilmark{4},
Michael~S.~Vogeley\altaffilmark{17},
David~H.~Weinberg\altaffilmark{18},
Brian~Yanny\altaffilmark{4}
}

\affil{%
{\small
${}^1$Institute for Advanced Study, Olden Lane, Princeton, NJ 08540\\
${}^2$Enrico Fermi Institute, University of Chicago, 5640 South Ellis Ave., Chicago, IL 60637\\
${}^3$Steward Observatory, 933 N. Cherry Ave., Tucson, AZ 85721\\
${}^4$Fermi National Accelerator Laboratory, P.O. Box 500, Batavia, IL 60510\\
${}^5$Princeton University Observatory, Princeton, NJ 08544\\
${}^{6}$Department of Physics and Astronomy, The Johns Hopkins University, 3701 San Martin Drive, Baltimore, MD 21218, USA\\
${}^{7}$Department of Physics and Astronomy, University of Pittsburgh, Pittsburgh, PA 15260\\
${}^{8}$Dept. of Physics, Carnegie Mellon University, 5000 Forbes Ave., Pittsburgh, PA 15232\\
${}^{9}$Yale University, P.O. Box 208101, New Haven, CT 06520\\
${}^{10}$Universidad de Chile, Casilla 36-D, Santiago, Chile \\
${}^{11}$Institute for Cosmic Ray Research, University of Tokyo, Midori, Tanashi, Tokyo 188-8502, Japan\\
${}^{12}$New York University, Department of Physics, 4 Washington Pl., New York, NY 10003\\
${}^{13}$Space Telescope Science Institute, 3700 San Martin Dr., Baltimore, MD 21218\\
${}^{14}$Palomar Observatory, 105-24, California Institute of Technology, Pasadena, CA 91125\\
${}^{15}$Physics Department, Rochester Institute of Technology, 85 Lomb Memorial Drive, Rochester, NY 14623-5603\\
${}^{16}$Department of Astronomy and Astrophysics, The Pennsylvania State University, University Park, PA 16802\\
${}^{17}$Department of Physics, Drexel University, Philadelphia, PA 19104\\
${}^{18}$Ohio State University, Dept.of Astronomy, 140  W. 18th Ave., Columbus, OH 43210\\
${}^{19}$Hubble Fellow\\[6pt]
}
{\it Accepted for the Astronomical Journal, November 2001 issue}
}
\setcounter{footnote}{19}
\begin{abstract}
We describe the target selection and resulting properties of a
spectroscopic sample of luminous, red galaxies (LRG)
from the imaging data of the Sloan Digital
Sky Survey (SDSS).  These galaxies are selected on the basis of color
and magnitude to yield a sample of luminous, intrinsically
red galaxies that extends fainter and further than the main flux-limited portion
of the SDSS galaxy spectroscopic sample.  The sample is designed to
impose a passively-evolving luminosity and rest-frame color cut to
a redshift of 0.38.  Additional, yet more luminous, red galaxies are
included to a redshift of $\sim\!0.5$.  Approximately 12 of these galaxies per
square degree are targeted for spectroscopy, so the sample will number
over 100,000 with the full survey.  SDSS commissioning data indicate
that the algorithm efficiently selects luminous ($M_{\gs}\approx-21.4$), 
red galaxies, that the spectroscopic success rate is very high, and that 
the resulting set of galaxies is approximately volume-limited out to $z=0.38$.  
When the SDSS is complete, the LRG spectroscopic sample will fill over 
$1\hgpcC$ with an approximately homogeneous population of galaxies and will 
therefore be well suited to studies of large-scale structure 
and clusters out to $z=0.5$.  
\end{abstract}

\section{Introduction}\label{sec:intro}
The Sloan Digital Sky Survey \citep[SDSS;][]{Yor00} combines a 
5-band CCD imaging survey
of the Northern Galactic Cap with an extensive and diverse multi-fiber
spectroscopic follow-up program.  The centerpiece of the spectroscopic survey
is a sample of 1 million galaxies.  This sample consists of
two parts.  The dominant portion, with about 88\% of the fiber allocation, is
a flux-limited sample (hereafter called MAIN) that will reach to
approximately $r\sim17.7$ \citep{Str01}.  
This sample has a median redshift of 
0.10 and few galaxies beyond $z=0.25$.  

The other 12\% of the galaxy spectroscopic sample is devoted to
galaxies that are fainter than the MAIN galaxy flux cut but expected, based on 
the observed colors, to be intrinsically red and at higher redshift.  
The strong 4000\AA\ break of early-type galaxies allows the SDSS
to acquire redshifts for these fainter galaxies in the same amount 
of observing time despite a lower signal-to-noise ratio.
At the outset, the goal of this luminous, red galaxy (LRG)
survey\footnote{Called bright, red galaxies (BRG), in analogy to
brightest cluster galaxies, in earlier papers and documentation.}
was to produce a volume-limited sample of intrinsically luminous ($\gtrsim\!3L^*$),
intrinsically red galaxies out to $z=0.5$.  The term ``volume-limited''
means that the same population of galaxies would be traced across
redshift.  In principle, evolution and merging make this a poorly-defined
concept.  However, luminous, red galaxies (e.g. giant ellipticals)
are observed to be evolving slowly 
\citep{Oke68,Sch71,Gun75,Rak95,Kau96,Lub96,Oke96,Ell97,Ara98,Col98,Sta98,van98,Bur00}, 
so we tune our selection 
to remove the passive evolution of an old stellar population.  Other modes
of evolution are left in the sample to be discovered.

A volume-limited sample of luminous, red galaxies is an efficient tool
for a number of important science goals.  First, because the brightest
galaxies in galaxy clusters tend to be very luminous and red 
\citep{San72,Hoe80,Sch83,Pos95}, 
the LRG sample will include
many luminous cluster galaxies and therefore will give spectroscopic
redshifts for clusters selected from the SDSS imaging survey 
\citep[e.g.,][]{Ann99,Nic00,Got01,Kim01}.  Second,
the sample will probe over $1\hgpcC$ with sufficient number density to yield
an immense volume for the study of large-scale structure.
Finally, the sample should permit studies of the evolution of giant
elliptical galaxies from $z=0$ to $z=0.5$.  The evolution of these
systems is an important, if controversial, probe of hierarchical 
galaxy formation \citep{Kau96b,Ara98,Col98}.

The purpose of this paper is to describe the selection algorithm used to
select the LRG sample and to assess how the resulting sample approaches
the design goals.  We leave most details of the SDSS hardware and data
reduction to other technical papers.  \citet{Yor00} provides an overview of
the survey.
The imaging data taken with the photometric camera \citep{Gun98}
through 5 filters \citep{Fuk96} 
is reduced with the software pipeline \photo\ \citep{Lup01a,Lup01b}.  
The photometric calibration is summarized in \citet{Sto01}.
The target selection occurs within
a software package known as \target\ 
using the algorithm described in this paper.  
The targets are then distributed onto an adaptive mesh of plates 
\citep{Bla01b}.  Spectra are obtained by a pair of 
fiber-fed double spectrographs \citep{Uom01,Cas01} and reduced by the 
\spectro\ software pipeline \citep{Fri01,Sch01}.
Further documentation of the survey can be found in the description
of the Early Data Release \citep{Sto01}.

The paper is arranged as follows.  
In \S \ref{sec:selection}, we describe the color-space context in which LRG 
selection occurs and then specify the cuts used to select the sample.
The minor differences between the current algorithm and the algorithms
used in the commissioning data are listed in Appendix \ref{sec:commissioning}.
We assess the photometric and spectroscopic performance of the 
sample in \S \ref{sec:performance}.  In appendix \ref{sec:appendix},
we specify the models used to generate $K$ and evolutionary corrections
for the data.  
We give advice and caveats about using the sample in \S \ref{sec:uselrg}
and conclude in \S \ref{sec:concl}.

We need to alert the reader to the fact that there are two different
usages of the term ``LRG sample'' in this paper.  On the one hand,
the question facing LRG target selection is 
how to choose the spectroscopic targets fainter than the MAIN sample flux limit.
Our assessment of sample efficiency, fiber quotas, and 
spectroscopic performance deals
only with these targets.  On the other hand, because LRGs are among
the most luminous galaxies, it is clear that at $z\lesssim0.3$ a
volume-limited sample of LRGs will include some galaxies that are
bright enough to be in the MAIN sample.  Hence, assessments of the
properties of the sample across redshift must include the LRGs from
the MAIN sample.  The choice of sample---LRGs with and without MAIN
sample contributions---should be clear from context. 

We have adopted a cosmology of $\Omega_m=0.33$ and $\Lambda=0.67$
for the calculation of distance moduli and comoving volumes.
All absolute magnitudes and comoving volumes are quoted assuming 
$H_0=100\kmsmpc$.

\section{Target Selection}
\label{sec:selection}
\subsection{Photometric Redshifts and LRGs}
Galaxies vary significantly in their luminosities and spectral energy distributions
(SEDs).  Fortunately, their SEDs are sufficiently regular that one can attempt
to infer their redshift from their colors \citep{Bau62,Koo85,Loh86,Con95}.
Were these photometric
redshifts completely accurate, one could apply luminosity and intrinsic
color cuts, and hence select the LRG sample, with fidelity.
However, the fact that the redshifts estimated from photometry have
errors complicates the isolation of the luminous, intrinsically red 
galaxies.

\begin{figure}[tb]
\plotone{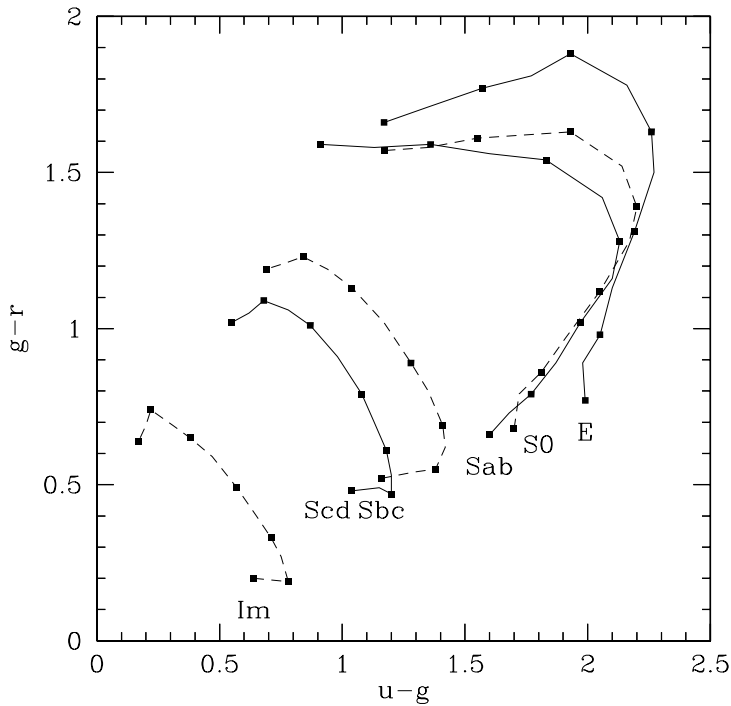} 
\caption{\label{fig:ugr}%
$u-g$ versus $g-r$ color for 6 non-evolving SEDs from 
\protect\citet{Fuk95} and \protect\citet{Ken92} as a progression
of redshifts.  Redshift $z=0$ is at the end near the label, each solid dot
represents an increment of 0.1 in redshift, and the last dots are $z=0.6$.  
The loci are reasonably separated in color-color space, indicating the 
possibility of accurate photometric redshifts.
}\end{figure}

\begin{figure}[tb]
\plotone{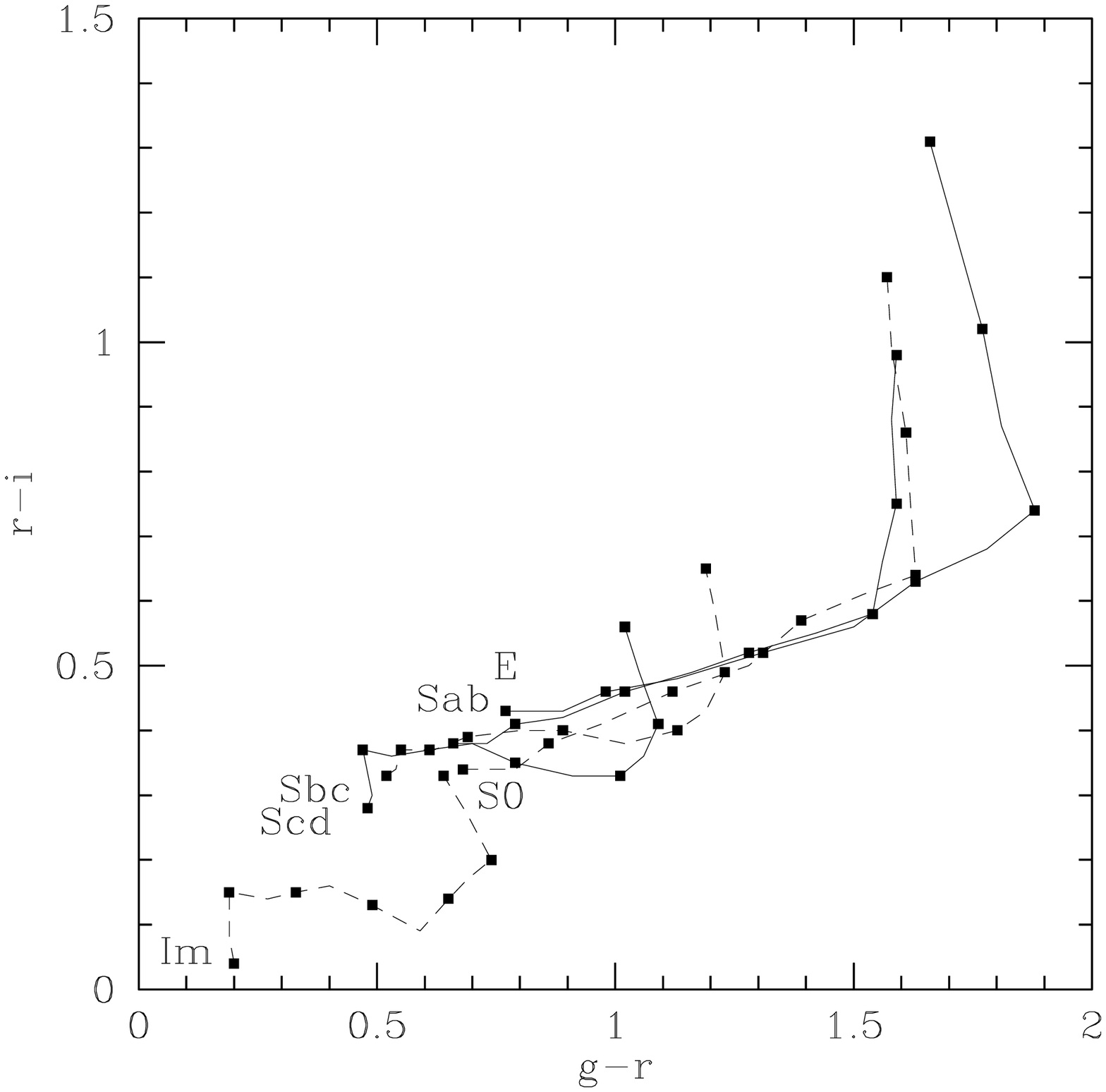} 
\caption{\label{fig:gri}%
As Figure \protect\ref{fig:ugr}, but for $g-r$ versus $r-i$.
Here, the redshift loci for different SEDs lie on top of one another,
indicating a photometric redshift degeneracy.
}\end{figure}

The errors in the photometric redshifts depend in detail upon the
bandpasses and sensitivity of the SDSS.  For an early-type galaxy, the
4000\AA\ break provides a sharp feature in the SED from which we can
infer the redshift.  For $z<0.38$, this feature lies within the SDSS $g$
band.  If all galaxies had the same SED, then the $g-r$ color 
would be an excellent redshift indicator.  However, since galaxies
actually show a range of 4000\AA\ break strengths, $g-r$ actually
measures only a degenerate combination of the position of the break
(i.e.~the redshift of the galaxy) and the strength of the break.
One can break this degeneracy with the $u-g$ color, as shown in 
Figure \ref{fig:ugr}.  With these two colors, one could infer both
the redshift and 4000\AA\ break strength of the galaxy, which would
allow selection of luminous, red galaxies.

Unfortunately, the galaxies of interest ($r\sim19$) have $u\sim22$, which is
close to the SDSS detection limit so that the measured $u-g$
color of these objects is quite noisy.  One must therefore
turn to the $r-i$ color, which is well-measured.  In principle, this 
second color would allow us to measure both redshift and SED type.
However, as shown in Figure \ref{fig:gri}, galaxy SEDs suffer from 
an accidental degeneracy in the SDSS bands at $z<0.4$, such that 
early-type SEDs of lower redshift have the same $g-r$ and $r-i$
color as later-type SEDs of a higher redshift.  In other words, 
galaxies of varying redshifts (less than 0.4) and SED form a nearly 
1-dimensional locus in $g-r-i$ space; we cannot infer two quantities
from the single position along this locus.  As we will see, the bivariate
luminosity-color distribution of galaxies does allow us to bypass this
problem.

At $z>0.38$, the situation improves as the 4000\AA\ break enters the
$r$ band.  The sensitivity of SDSS in $g$ is sufficient 
at $r\sim19.5$ to yield a well-measured $g-r$ color, and one can combine
this with $r-i$ to constrain the redshift independently of the SED
of the galaxy.  This can be seen in Figure \ref{fig:gri} from the fact that
the curves for the different SEDs do not overlap at $z>0.4$.

Because of this transition at $z\approx0.4$, we are driven to
use different selection cuts at $z\lesssim0.4$ and $z\gtrsim0.4$.
At $z\lesssim0.4$, we must extract LRGs out of the galaxy locus by relying
on the bivariate galaxy luminosity and rest-frame color distribution
to discriminate against bluer or less luminous galaxies.  
Similar selections have been performed by \citet{War93}, \citet{Gla00}, and 
\citet{Wil01}.
At $z\gtrsim0.4$,
it is easy to isolate the LRGs, although we will find that the finite
spectroscopic integration time keeps us from fully extending to the desired
luminosity cut.

\subsection{SDSS Photometry}

Consistency tests within the SDSS data 
indicate that the relative photometry of the survey is very good;
however, the bandpasses of the filters are measured to be slightly
different than designed \citep{Fuk96,Fan01,Sto01}, 
and the zero-point of the magnitude
systems, which were intended to satisfy the ${\rm AB}_{95}$ convention,
are still provisional.  Hence, instead of referring 
to the magnitudes as $u$, $g$, $r$, $i$, and $z$, we refer
to the current photometric solutions, as codified in the Early
Data Release \citep{Sto01}, as $\us$, $\gs$, $\rs$, $\is$, and $\zs$.
This is a minor problem as regards the selection of LRGs, since 
the final calibration will only change the zeropoints of the magnitude
system and hence move the selection cuts in easily calculable ways.  
However, at the time of this writing, the uncertainties over the
bandpass shape and zeropoints create difficulties regarding the 
{\it interpretation} of the LRG spectroscopy.  Because the sample spans a 
range of redshift, one must model the time-evolving SED of the galaxies in
order to make corrections for the redshifting
of the bandpasses and the evolution of the stellar populations.
Those models do not 
produce colors that match our photometry, presumably in part because of some
remaining problems in our knowledge of the filter shapes and zeropoints.
The models and the corrections we apply
are described in Appendix \ref{sec:appendix}.

\subsection{Selection Cuts}
\label{sec:cuts}

As described above, we use different techniques above and below $z\approx0.4$.
The low-redshift cut, which accounts for 80-85\% of the targets, will
be called Cut I; the high-redshift cut will be called Cut II.
As a matter of context, we note that the MAIN sample flux limit is
$\rs\sim17.7$, whereas the commissioning spectroscopic data
show that a 45-minute exposure with the SDSS 2.5m telescope can
acquire reliable redshifts on objects with strong 4000\AA\ breaks
as faint as $\rs\sim19.5$.  Hence, the LRG cuts are optimized to
work in this range of magnitudes.

For both cuts, we rely on \citet{Pet76} magnitudes in the $r$ band to set our
flux and surface brightness cuts.  
These magnitudes are calculated exactly as for the MAIN galaxy
sample \citep{Str01}, avoiding any discontinuity in 
the transition from the MAIN to LRG samples.
The $\rs$ surface brightness is calculated using the radius inside which
half the Petrosian flux is found, i.e.,
\beq\label{eq:defSB}
\sbr = \rpet + 2.5 \log_{10}(2\pi R_{50}^2).
\eeq
Again, this choice matches that of MAIN galaxy target selection.

For colors, we use the model magnitudes from \photo.
The best-fit exponential or de Vaucouleurs model, allowing for arbitrary scale
length and axial ratio and convolving with the local point spread function,
is found for each object in the $r$ band.  
That model is then used to extract the flux
in the other bands.  Since all bands are measured with the same effective
aperture, the colors are unbiased in the absence of color gradients;  
the resulting colors have higher signal-to-noise ratio
than a simple aperture color.  All colors in LRG selection refer to
differences of model magnitudes.  Further details of model magnitudes 
can be found in \citet{Lup01b}.  Finally, the separation of stars from
galaxies is done by differencing the $\rs$ model magnitude from the
$\rs$ PSF magnitude, as described in \citet{Lup01b}.  This is the same
method as for MAIN galaxy target selection, but we will use a 
different value for the threshold parameter.

All magnitudes and colors have been corrected for Galactic extinction
using the \citet{Sch98} map and assuming $R_V=3.1$.

Galaxy colors in the $\gs-\rs$ vs. $\rs-\is$ plane display
a narrow linear locus due to the degeneracy of early-
to mid-type galaxies at $z<0.4$.  We therefore adopt a rotated
coordinate system in color-space so that we can measure position along and across
that locus.  The distance perpendicular to the locus is simply
\beq\label{eq:defcperp}
\cperp = (\rs-\is) - (\gs-\rs)/4.0 - 0.18.
\eeq
The intercept is chosen from the data so that $\cperp=0$ marks the center of
the observed locus.

For galaxies that lie in the locus, we wish to estimate where they
fall along the locus.  One could use the orthogonal\footnote{Note that
this ``orthogonality'' is a notational illusion; there is no physical motivation for a
Euclidean metric in the $\gs-\rs$ vs. $\rs-\is$ plane.} complement of
equation (\ref{eq:defcperp}), but if we assume that galaxies
fall along the line $\cperp=0$, we are free to choose the linear
combination of colors that minimizes the error in the estimator.  In
other words, because the error of $\gs-\rs$ is different from that of
$\rs-\is$, one may get a more accurate estimate of the position of 
the galaxy along the locus by weighing one color more than the other.
Given the average errors on these three magnitudes for galaxies in the
given flux and color range, we adopt
\beq\label{eq:defcpar}
\cpar = \cparcoeff (\gs-\rs) + (1-\cparcoeff) 4.0 [(\rs-\is)-0.18].
\eeq
The definition is chosen so that an object with
$\cperp=0$ has $\cpar=\gs-\rs$.  We use $\cparcoeff=0.7$.  
In principle, one could have chosen
$\cparcoeff$ on an object-by-object basis so as to minimize the 
error on $\cpar$ given the quoted errors on the colors.  However, 
this would have reduced the formal errors only slightly at the 
expense of requiring the formulae to depend sensitively on the
quoted errors.

In fact, the error on $\cpar$ is insensitive to the choice of $C$, but our choice
does help in a different fashion.  Although the early-type color-color locus
makes a sharp turn at $z\approx0.4$, the galaxies do not move out of the
$z<0.4$ locus immediately, simply because of the non-zero thickness of
that locus.  By choosing $\cparcoeff=0.7$ as opposed to 16/17 (which would
have established a strictly orthogonal system), galaxies at $z>0.4$
still have $\cpar$ grow with redshift.  This procedure makes Cut I effective to
slightly higher redshifts than one would have guessed {\it a priori}.

Objects that are flagged by \photo\ as BRIGHT or SATURated in $\gs$, $\rs$,
or $\is$ are excluded.  Objects must be detected as BINNED1, BINNED2,
or BINNED4 in both $\rs$ and $\is$, but not necessarily in $\gs$.
These flags are explained in \citet{Sto01}.  The BINNED flags indicate
that the object is detected in the particular band, where detected means
that one or more (original or $2\times2$ or $4\times4$ re-binned) pixels 
is more than 5-$\sigma$ above the sky level.  Since the faintest LRG
candidates are quite faint in $\gs$, we do not wish to require such 
a detection in that band; however, 99\% of the targets are in fact so detected in $\gs$.
As with all SDSS target selection, objects are considered after
deblending, with each fragment judged separately.

LRG target selection in the commissioning data \citep{Sto01} used slightly
different cuts than those described below.  The differences are listed in
Appendix \ref{sec:commissioning}.

\subsubsection{Cut I ($z\lesssim0.4$)}

\begin{figure}[tb]
\plotone{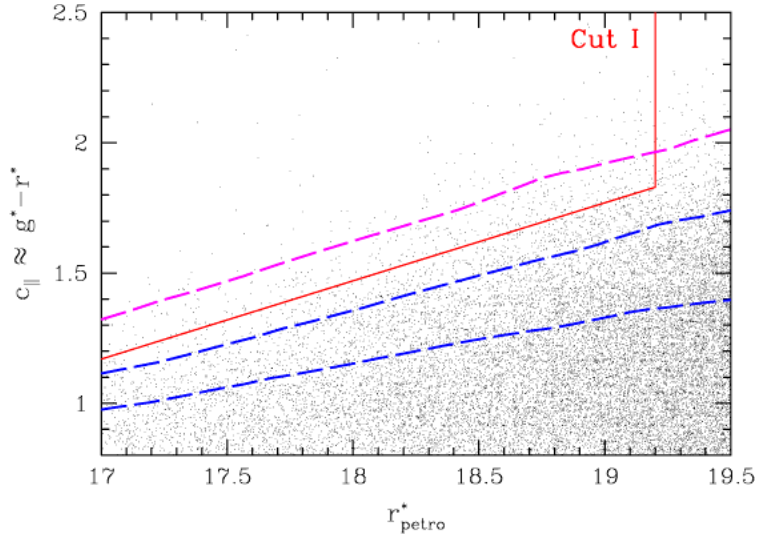} 
\caption{\label{fig:rpetcpar}%
Petrosian $\rs$ apparent magnitude versus observed color $\cpar$ for a 
set of galaxies from SDSS.  The solid lines show the selection region
for Cut I LRGs.  The dashed lines show three loci predicted by a stellar
population synthesis model for galaxies as a function of redshift.  
The top line is for a passively-evolving old population 
(appendix \protect\ref{sec:appendix}); the lower
two lines mix in progressively more late-time star formation.  
Of course, changing the absolute magnitude of a galaxy will shift the
lines horizontally; the displayed lines have $z=0$ $\rs$ absolute magnitudes of
$-22.2$, $-21.7$, and $-21.7$, top to bottom.  We
use the fact that the old population has a nearly linear magnitude-color
relation in our selection cut.  The data are taken from
SDSS imaging runs 752 and 756 with $185^\circ<\alpha<235^\circ$ and $|\delta|<1.25^\circ$,
excluding a few fields from run 752 with $\rs$ seeing FWHM worse than
$2''$ \protect\citep{Sto01}.
}\end{figure}

\begin{figure}[tb]
\plotone{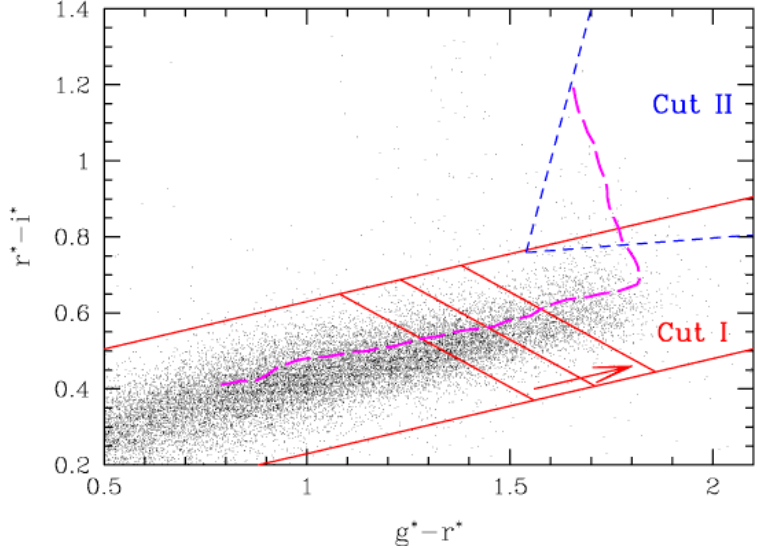} 
\caption{\label{fig:grri}%
The $\gs-\rs$ versus $\rs-\is$ color-color diagram for galaxies
with $18.5<\rs<19.5$ from SDSS.  The red solid lines show the selection region
for Cut I LRGs.  The three lines overlaid with an
arrow indicates that the location of the line cutting across
the galaxy locus is a function of $\rs$ apparent magnitude; fainter galaxies
must be redder to pass the cut.  
The displayed lines correspond to $\rs=17.5$, 18.0, and 18.5, left to right.
The blue short-dashed lines show the (magnitude-independent)
selection region for Cut II LRGs.  The long-dashed line shows the locus
of a passively-evolving old population as a function of redshift
(appendix \protect\ref{sec:appendix}); the 
bend in the locus occurs at $z\approx0.40$.
The galaxy sample is the same as in Figure \protect\ref{fig:rpetcpar}.
}\end{figure}

\begin{figure}[tb]
\plotone{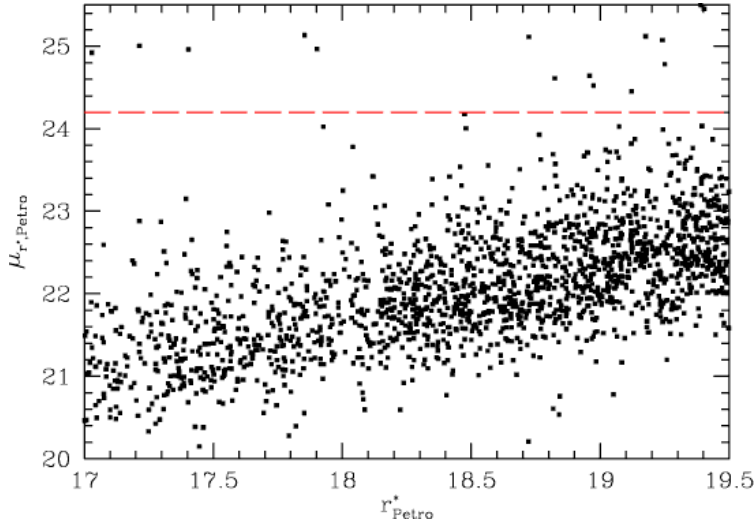} 
\caption{\label{fig:rpetsbr}%
Petrosian $\rs$ apparent magnitude versus Petrosian $\rs$ half-light
surface brightness $\sbr$ for 1700 galaxies that would otherwise
pass either of the LRG cuts.  
Only the flux cut (eq.~[\protect\ref{eq:cutIrpet}])
and surface brightness cuts (eqs.~[\protect\ref{eq:cutIsbr}] and 
[\protect\ref{eq:cutIIsbr}]) have been omitted.  One sees that the 
surface brightness cut, indicated by the dashed line, eliminates only a 
small fraction of potential targets.  A number of junk objects, 
e.g.~scattered light, occur at $\sbr>26$.
The data from which these objects were selected is as in Figure \protect\ref{fig:rpetcpar}.
}\end{figure}

\begin{figure}[tbp]
\plotone{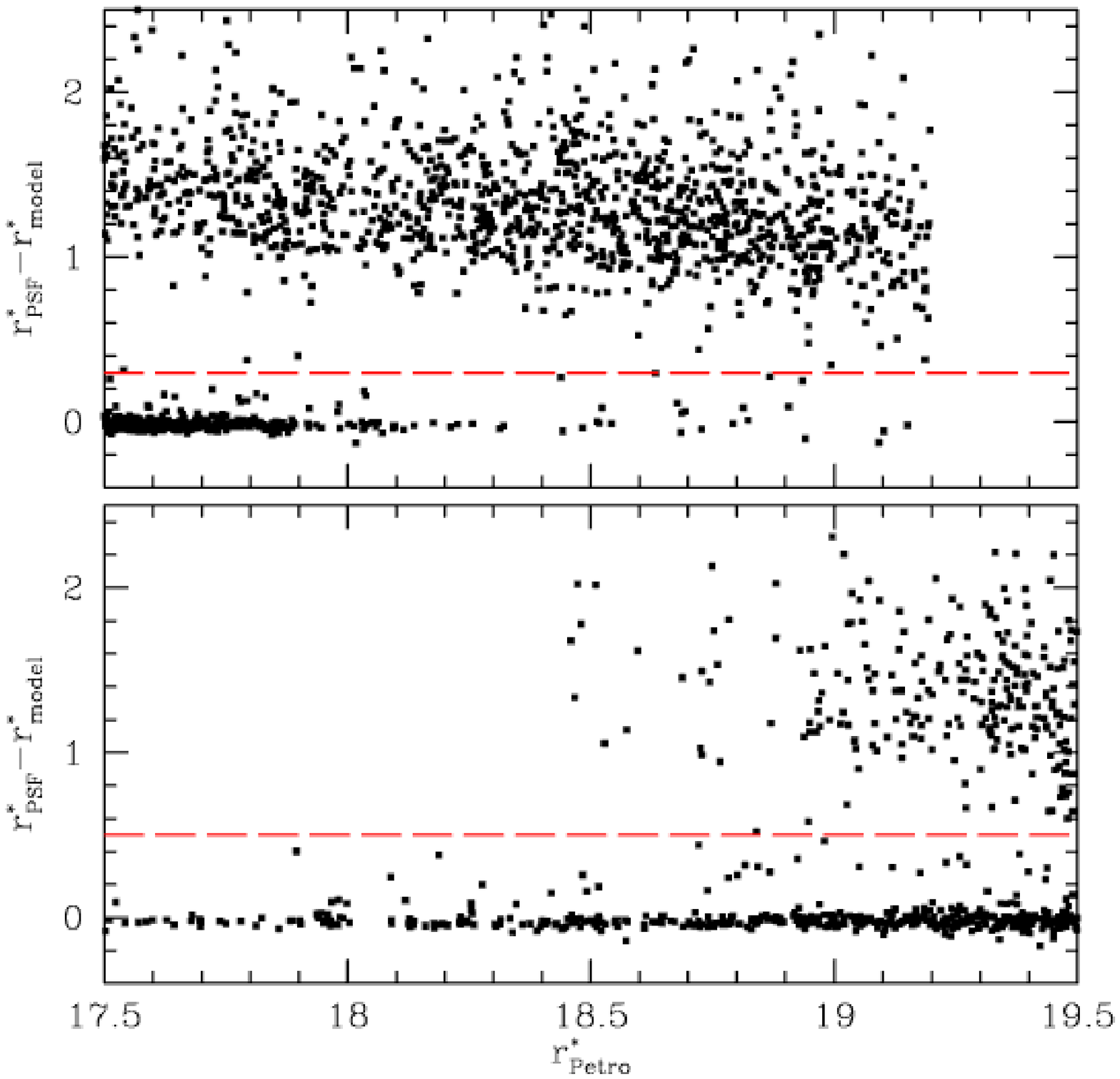} 
\caption{\label{fig:rpetrdiff}%
Petrosian $\rs$ apparent magnitude versus $\rpsf-\rmodel$, the quantity
used for star-galaxy separation.  
({\it top panel}) All objects that pass Cut I excluding the star-galaxy
separation cut.  Fainter magnitudes have fewer stars because the color
cuts have moved away from the stellar locus.  
({\it bottom panel}) Same, but for Cut II.  One sees
that the ratio of stars to galaxies is less favorable for Cut II, hence
the more restrictive threshold of 0.5 rather than 0.3.  
Of the few stars discovered spectroscopically in the sample above the
threshold, many fall well above the cut.
Undeblended close pairs of stars as well
as faint stars with superposed diffraction spikes from bright companions can
boost $\rmodel$ significantly brightward of $\rpsf$.
The region of the data is as in Figure \protect\ref{fig:rpetcpar}.
}\end{figure}

Cut I uses $\cpar$ as a redshift indicator and uses a sliding flux cut
so as to approach a constant passively-evolving luminosity cut.  We
consider the locus of $\rs$ vs. $\cpar$ for an old single-age stellar 
population.  For all models considered, this locus is nearly linear
across the redshift range of interest, as shown in Figure \ref{fig:rpetcpar}.  
Therefore, we select objects
that satisfy
\beqa\label{eq:cutI}
\rpet &<& 13.1 + \cpar/0.3, \\
\rpet &<& 19.2, \label{eq:cutIrpet}\\
|\cperp| &<& 0.2, \\
\sbr &<& 24.2 {\rm\ mag\ arcsec^{-2}}, \label{eq:cutIsbr}\\
\rpsf - \rmodel &>& 0.3. \label{eq:cutIsg}
\eeqa
Equation (\ref{eq:cutI}) is the most important, because it is primarily
responsible for setting the luminosity threshold as a function of redshift.  
Equation (\ref{eq:cutIrpet}) imposes a flux cut chosen in large part 
to ensure good spectroscopic performance.
The $\cperp$ cut restricts attention to the $z<0.4$ galaxy locus
(see Figure \ref{fig:grri}).
The $\sbr$ cut removes very low surface brightness objects that 
are often reduction errors of various sorts, e.g. deblended diffraction
spikes or scattered light \citep{Str01}.  This cut is
important because such objects can have strange colors.  As shown
in Figure \ref{fig:rpetsbr}, the surface brightness threshold is
safely fainter than most LRG candidates.   The
$\rpsf-\rmodel$ cut is our star-galaxy separator.  Again, as
shown in Figure \ref{fig:rpetrdiff}a, most Cut I LRGs are
safely away from this cut.

As additional quality assurance cuts, we exclude objects that have
$\gs-\rs>2.5$, $\rs-\is>1.5$, or estimated errors on the model magnitudes exceeding
0.2, 0.1, or 0.1 mag in $\gs$, $\rs$, and $\is$, respectively.  Very few
objects fail any of these cuts without also failing the $\sbr$ cut.

The number of objects selected is extremely sensitive to the
color-magnitude cut in equation (\ref{eq:cutI}).
A shift of 0.01 in $\cpar$ or 0.03 in $\rpet$
makes a 10\% change to the number of galaxies selected.
We will discuss this further in \S\ \ref{sec:caveats}.

As we discussed above, $\cpar$ alone is not sufficient to predict the redshift of
a galaxy.  Intrinsically bluer galaxies at higher redshifts or intrinsically
redder galaxies at lower redshifts can produce
the same colors (Fig.~\ref{fig:gri}).  
However, Cut I turns out to be extremely effective at
selecting LRGs.  The reason is the shape of the bivariate 
luminosity-color function,
not the success of photometric redshifts.  Intrinsically redder
galaxies at lower redshift---and hence lower luminosity at a given 
flux---would satisfy the cut, but there are few luminous
galaxies redder than the old stellar populations of giant elliptical and
cD galaxies.  Conversely, intrinsically bluer galaxies at higher redshift
and higher luminosity would pass the cut, but such super-luminous galaxies
are extremely rare.  Hence, as will be shown later, most of the galaxies
that pass the cut turn out to have redshifts that place their intrinsic
colors and luminosities in the ranges appropriate for old stellar systems
well above $L^*$.

Objects that satisfy both the LRG cut and the MAIN cut are flagged as
both by the \target\ pipeline.
However, the linear color-magnitude cut 
(eq.~[\ref{eq:cutI}]) is not a good approximation to the locus
of an early-type galaxy at lower redshifts.  At $z<0.15$, Cut I
is too permissive, allowing lower luminosity sources to enter the
LRG sample.  Hence, to extract LRGs from the MAIN sample at $z<0.15$, 
one must make additional post-spectroscopic cuts.  These are described
in \S\ \ref{sec:main}.

\subsubsection{Cut II ($z\gtrsim0.4$)}

Cut II is used to select LRGs at $z>0.4$ by identifying galaxies that
have left the low-redshift locus
in the $\gs-\rs$ vs. $\rs-\is$ plane.  At these redshifts, 
we can distinguish 4000\AA\ break
strength from redshift, so we can isolate intrinsically red galaxies.
The difficulty is avoiding interlopers, either from $z\lesssim 0.4$
galaxies that scatter up in color from the low-redshift locus or 
from late-type stars, which are far more numerous.

We adopt $\rpet=19.5$ as our flux limit because fainter objects would 
not reliably yield sufficient signal-to-noise ratio in the spectra.
Unfortunately, the luminosity threshold in Cut I would predict 
$\rpet>19.5$ at the redshifts of interest in Cut II.  Therefore, 
Cut II is simply a flux-limited sample with no attempt to produce
a fixed luminosity cut across the (narrow) range of redshift probed.

The selection imposed is
\beqa\label{eq:cutIIrpet}
\rpet &<& 19.5, \\
\cperp &>& 0.45 - (\gs-\rs)/6, \label{eq:cutIIcperp}\\
\gs-\rs &>& 1.30 + 0.25 (\rs-\is). \label{eq:cutIIgr}\\
\sbr &<& 24.2, \label{eq:cutIIsbr}\\
\rpsf - \rmodel &>& 0.5, \label{eq:cutIIsg}
\eeqa
Equation (\ref{eq:cutIIcperp}) separates the high-redshift region 
from the low-redshift galaxy locus.  It is tilted in $\cperp$ to
guard against the larger areal number density of prospective
interlopers at lower $\gs-\rs$.  Equation (\ref{eq:cutIIgr}) isolates
intrinsically red galaxies and separates the selected region from
the bulk of the late-type stellar locus.  These cuts are
displayed in Figure \ref{fig:grri}.  
Equation (\ref{eq:cutIIsbr})
places the same surface brightness cut as in cut I.  As before, this
is primarily a quality assurance cut.  
Finally, equation (\ref{eq:cutIIsg})
implements the star-galaxy separation; because of the larger ratio
of stellar objects to galaxies in the color-magnitude region, we 
adopt a stronger cut than in Cut I (see Figure \ref{fig:rpetrdiff}b).  
We again exclude objects with
$\gs-\rs>2.5$, $\rs-\is>1.5$, or errors on the model magnitudes exceeding
0.2, 0.1, or 0.1 mag in $\gs$, $\rs$, and $\is$, respectively.  Note that
the $\rs-\is$ cut here does exclude a few very late-type stars, whereas
early-type galaxies are not predicted to get this red, even for $z>0.6$.

The threshold of $\cperp$ needed to keep away from the
low-redshift locus restricts Cut II to $z\gtrsim0.43$.  Fortunately,
Cut I includes plenty of galaxies in the range $0.38 < z < 0.44$ because
of the definition of $\cpar$.

\section{Performance in Commissioning Data}
\label{sec:performance}

Many thousand spectra of LRGs have been acquired by the SDSS thus far.
The SDSS uses a pair of fiber-fed double spectrographs to cover the wavelength range 
from 3800\AA\ to 9200\AA\ with a resolution of $\sim\!1800$ ($\sim\!150\kms$ FWHM).
The diameter of the fibers are $3''$.
These spectra allow a detailed study of the properties of the objects
selected by the LRG cut.  

\begin{figure}[p]
\plotone{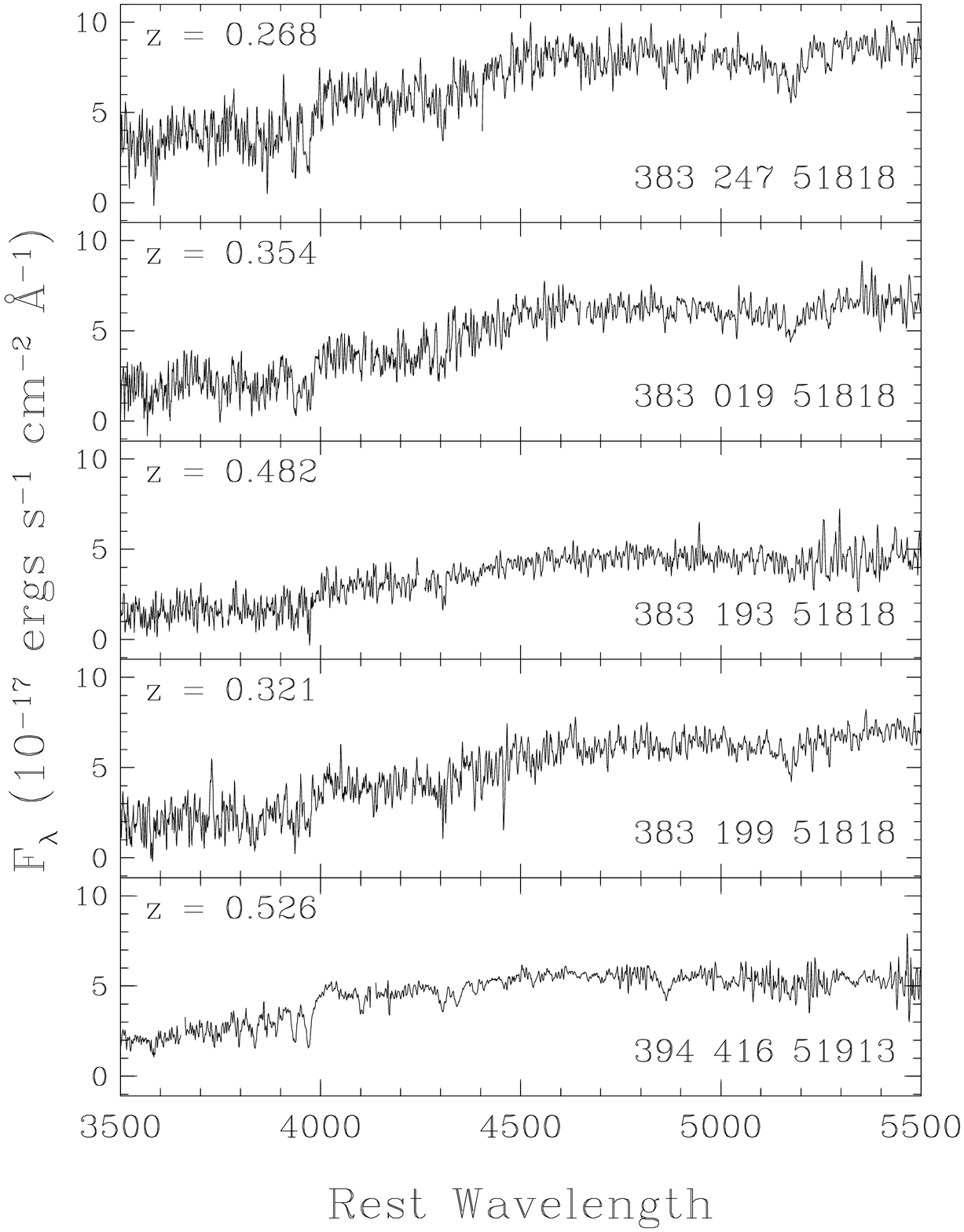} 
\caption{\label{fig:montage}%
Spectra of 5 LRG, with the wavelengths shifted to the rest-frame.
The top four spectra are taken from plate 383 observed on MJD 51818.  This
plate exceeded the survey minimum signal-to-noise ratio by about 20\%,
which is fairly typical.  The top 3 spectra (fibers 247, 19, and 193)
show a progression of redshift, while the 4th (fiber 199) is an unusual 
spectrum showing [\ion{O}{2}]~$\lambda 3727$ emission.  
The last spectrum is fiber 416 from plate 394 observed on MJD 51913.
We obtained much more signal on this plate, about 2.5 times the survey
minimum signal-to-noise ratio.  This spectrum shows strong Balmer absorption
characteristic of E+A galaxy spectra.  
The spectra have been slightly smoothed for presentation.
}
\end{figure}

A sample of LRG spectra are displayed in Figure \ref{fig:montage}.
The signal-to-noise ratio in the first four spectra are typical of the
survey; the bottom spectrum is well above the usual survey quality.
The first three spectra illustrate a progression of redshift, the
fourth spectrum shows a mild emission line, and the last spectrum
shows extra Balmer absorption characteristic of E+A galaxy spectra
\citep{Dre83}.  The latter two phenomena occur only rarely in the sample.

For a careful determination of the redshift distribution and failure
modes, we focus on a sample of 21 plates.  These were picked to be
in chunks 8, 9, 10, and 11 so that the final selection could
be implemented exactly (see Appendix \ref{sec:commissioning} for
an explanation of how targets were selected in the commissioning
data as well as a definition of the term ``chunk'').  
14 of these plates have duplicate observations
of survey quality.  Since the highest signal-to-noise ratio example of each
plate was picked, the primary set of 21 tends to have signal-to-noise ratios
significantly above the survey minimum.  These plates have 972 objects
selected by Cut I and 165 selected by Cut II (all fainter than the MAIN
survey).  We inspected all of these spectra by eye
to validate the redshift assigned by the \spectro\ pipeline.  A few
mistakes were fixed, and a few unknown cases were determined by eye.
The result is a very complete and accurate redshift catalog.

For the measurement of quantities where large-scale structure 
is the dominant uncertainty, we use
a larger sample of plates.  
This includes 8267 Cut I and 1284 Cut II LRG targets on 173 plates 
from Chunks 4-13.  In cases where an object
was observed multiple times, either because the plate was observed twice
or because of quality-assurance fiber allocation, the highest 
signal-to-noise ratio spectrum was used 
(as measured by the median signal-to-noise
ratio of the plate).  Some figures use a randomly selected
subsample of these plates, simply to reduce the density of points.

\subsection{Redshifts and derived quantities}

We begin by describing the results of the spectroscopy, focusing on
the set of 21 verified plates.  Of the 972 Cut I LRG targets on these plates,
7 were stars.  The remainder had redshifts between 0.10 and 0.53, of which
only 4 were at $z<0.2$.  None of the spectra failed to yield a redshift when 
examined by eye, although data problems
(e.g.~large patches of missing data because of bad CCD columns)
made a few cases difficult.  Nearly all were early-type galaxy spectra,
but some have emission lines in addition.

\begin{figure}[tb]
\plotone{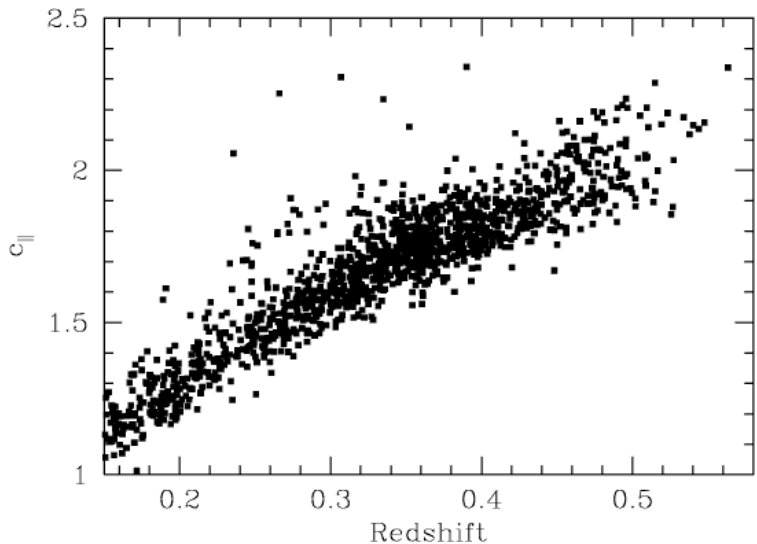} 
\caption{\label{fig:zzcpar}%
Redshift versus $\cpar$ color for LRGs, including objects from 
the MAIN sample that pass an LRG cut (always Cut I) and have $z>0.15$.  
}\end{figure}

Figure \ref{fig:zzcpar} plots redshift versus $\cpar$ color.  
Clearly, most galaxies fall along a tight correlation.
Excluding 4\% of the points as outliers, the {\it rms} scatter in $\cpar$ 
from a linear fit 
is $\sim\!0.065$ mag at $z<0.35$ and $\sim\!0.08$ mag at
$z>0.35$.  The scatter in redshift is 0.02 at $z<0.35$ and 0.03
at $z>0.35$.
A few percent of objects have redshifts that are low for an old stellar
population of their color; turning this around, one would say that their
rest-frame colors are redder than an old population given their redshift.
The most extreme example of this is an edge-on disk galaxy at $z=0.108$,
in which dust extinction is presumably strongly reddening the colors.
Close binary pairs of galaxies are another form of outlier; 
the deblending algorithm \citep{Lup01b} can occasionally deblend the different
bands inconsistently
(this behavior will be improved with future versions of the algorithm).  
If one is
interested in a strict sample of luminous galaxies with old stellar
populations, one should reject objects with rest-frame colors that are
too red.  Even if the color became too red because of an error in the
photometry, the
red claimed color means that the apparent magnitude cut is too permissive.
The result is that nearly all these objects are far less luminous than
one intended, as shown in the absolute magnitude versus rest-frame color 
plot in Figure \ref{fig:Mgug0cutI}.
As described in Appendix \ref{sec:appendix},
we use the measured redshift and 
the observed $\rpet$ magnitude and $\gs-\is$ color to 
construct the rest-frame, passively-evolved $\gpet$ absolute magnitude
and $\us-\gs$ color.
In this sample, only 7 objects (including the edge-on disk
mentioned above) have rest-frame $\us-\gs>2.35$ and 41 more have 
$2.1<\us-\gs<2.35$, as compared to a typical $\us-\gs$ of around 1.9.
No object is significantly too blue for its redshift.  Conservatively,
then, one would call the sample 95\% efficient at selecting luminous,
red galaxies.

\begin{figure}[tb]
\plotone{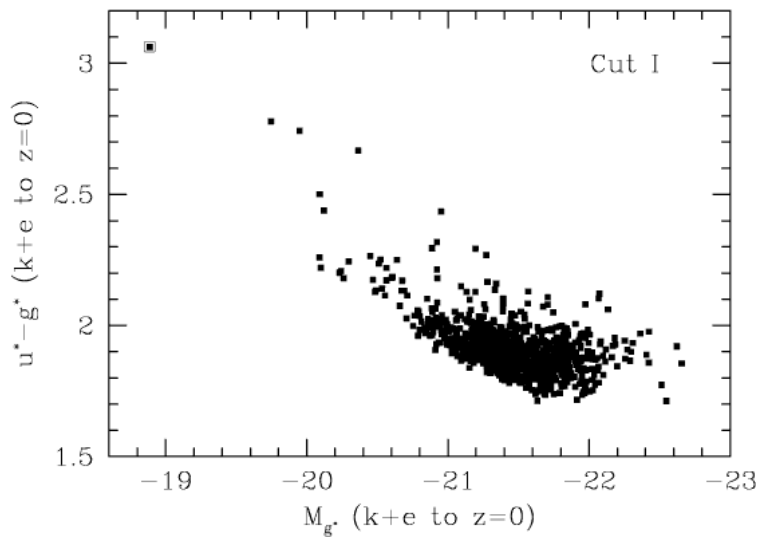} 
\caption{\label{fig:Mgug0cutI}%
Absolute magnitude versus rest-frame color for 965 Cut I LRGs
from our sample of 21 verified plates.  Both quantities have been 
$K$-corrected and passively-evolved to $z=0$ using the prescription
in Appendix \protect\ref{sec:appendix}.  
The sharp diagonal boundary in rest-frame $\us-\gs$ versus $M_{\gs}$ is
the result of the color-magnitude selection in equation (\ref{eq:cutI}).
The edge-on disk galaxy 
at $z=0.108$ is marked ({\it dot in open square})
as an extreme example of how an object much redder
than an old stellar population can enter the sample.}
\end{figure}

\begin{figure}[tb]
\plotone{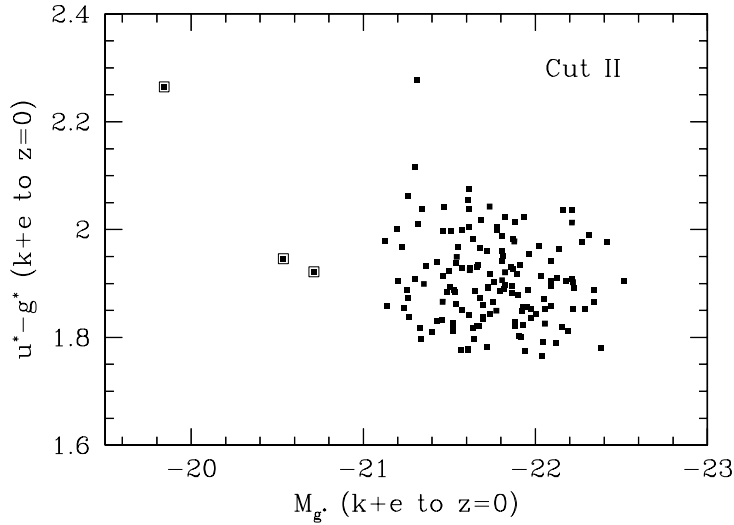} 
\caption{\label{fig:Mgug0cutII}%
Absolute magnitude versus rest-frame color for 150 Cut II LRGs
from our sample of 21 verified plates.  Both quantities have been 
$K$-corrected and passively-evolved to $z=0$.  Galaxies at $z<0.4$
have been marked as dots in open squares; 
the object at $z=0.08$ falls off the plot 
faintward and blueward.
Note that the bounding region of the cut is
not sloped as for Cut I (see Fig.~\protect\ref{fig:Mgug0cutI}).
}\end{figure}

\begin{figure}[tb]
\plotone{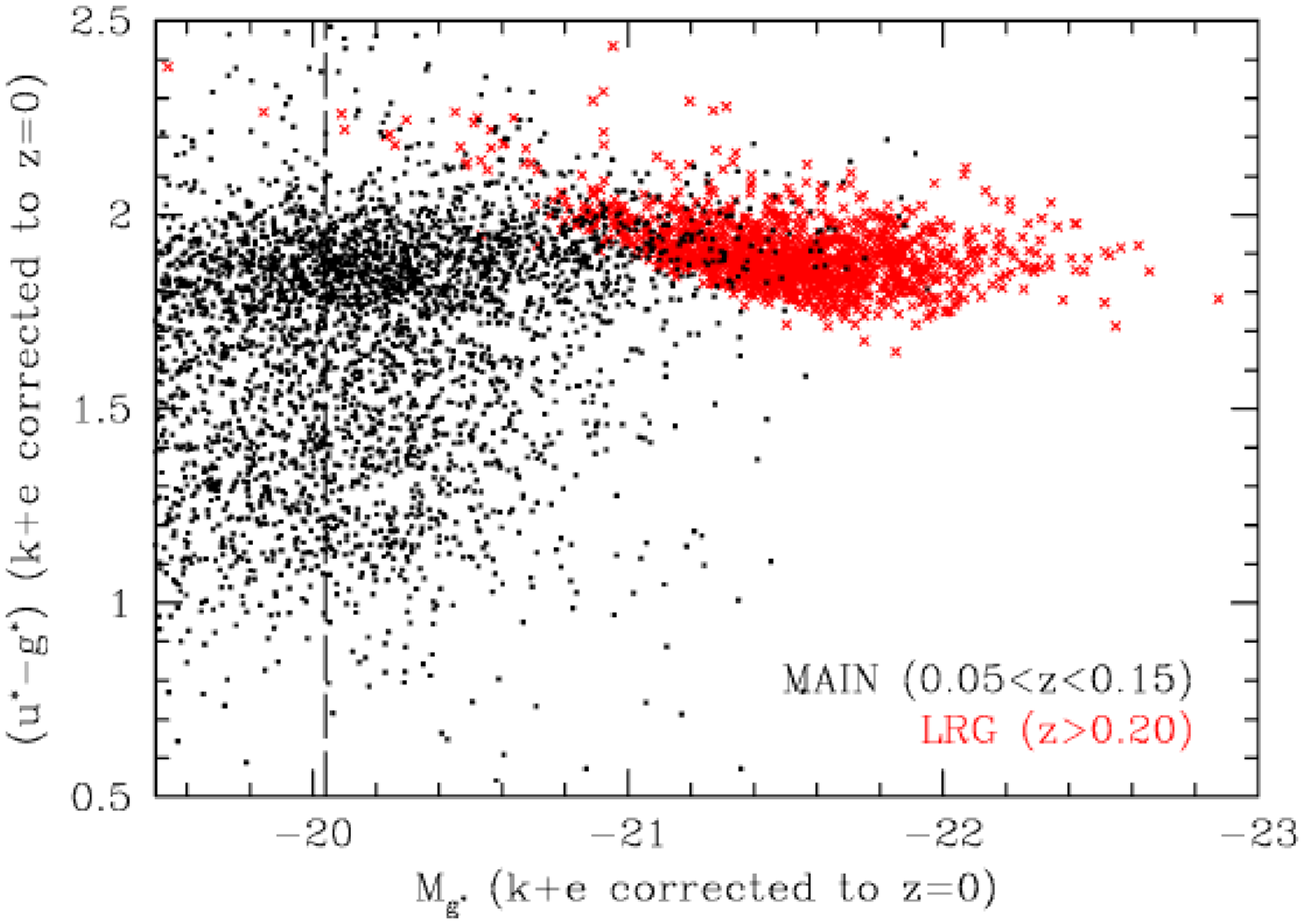} 
\caption{\label{fig:Mgug0use}%
Comparison of the absolute magnitude and rest-frame color distribution
of MAIN galaxies ({\it dots}) and LRGs ({\it crosses}) 
from our sample of 21 verified plates.  
The quantities on both axes have been 
$K$-corrected and passively-evolved to $z=0$.  
Only MAIN galaxies with redshifts between 0.05 and 0.15 and
LRGs at $z>0.2$ (including those from the MAIN sample) 
are shown.  The dashed line shows the
rough position of $M^*$ from the luminosity function of MAIN
galaxies \protect\citep{Bla01a}.  Additionally, the MAIN sample is
approximately volume-limited to the right of the dashed line.
One sees that the LRGs populate the luminous end of the red sequence.
The offset in the color of the red sequence between the MAIN and LRG
samples is presumably due to errors in the photometric zeropoints and/or
errors in the broadband shapes of the stellar population synthesis models
used to correct colors to zero redshift.
}
\end{figure}

Of the 165 Cut II LRGs, 11 were stars.  Four others had spectra with too
little signal to find a secure redshift; it turns out that one
of these was a deblended diffraction spike from a bright star.  
Of the objects with confirmed $z>0$ redshifts,
one is a $z=0.08$
galaxy with a late-type star superposed and three others have $z<0.38$
and fall outside of the $M_{\rs}$ vs. $\gs-\rs$ region populated by Cut I 
galaxies.  
The remaining 146 have redshifts between 0.38 and 0.57; all but five have
$z>0.42$.  Combining the 11 stars with the four lower-redshift galaxies
implies a success rate of finding luminous, red galaxies of about 90\%.

Figure \ref{fig:Mgug0cutII} shows the distribution of absolute
magnitudes and rest-frame colors for Cut II galaxies.  Comparing
this to the results for Cut I in Figure \ref{fig:Mgug0cutI} reveals
significant differences between the two samples.  Cut II achieves
a threshold in luminosity that is approximately independent of rest-frame
color, whereas Cut I has a strong correlation between the two.
In both cases, however, the selected galaxies do occupy the luminous,
red tail of the galaxy distribution.  This is shown in
Figure \ref{fig:Mgug0use}, which overplots the LRG sample against
a volume-limited sample of MAIN galaxies.

\subsection{Photometric Properties}
\label{sec:magnitudes}

\begin{figure}[tb]
\plotone{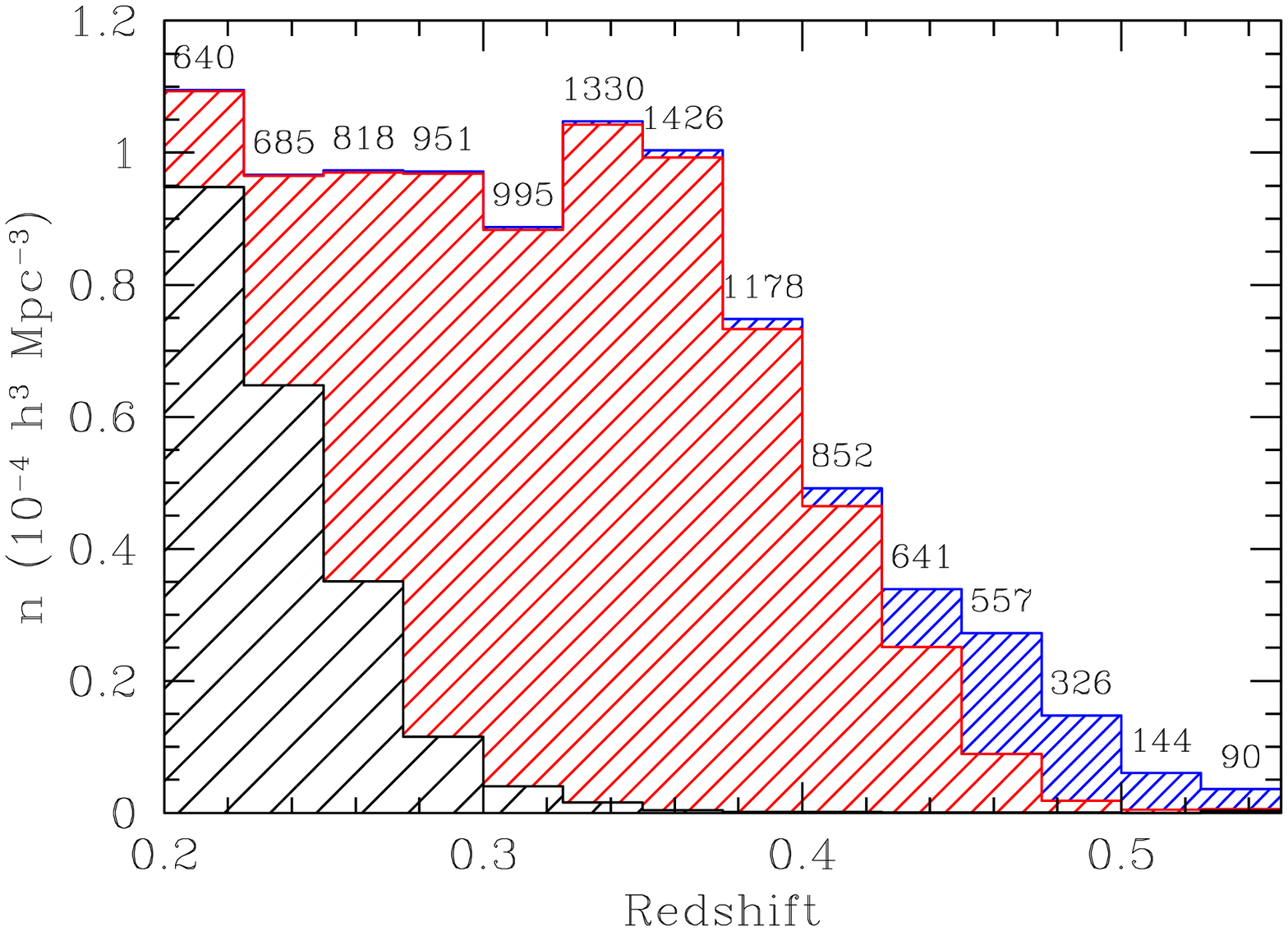} 
\caption{\label{fig:zzhist}%
The comoving number density of LRGs as a function of redshift.
The shaded regions from left to right indicate LRGs from the MAIN sample, 
Cut I (fainter than the MAIN flux limit), and Cut II
contributions.  The numbers indicate the number of galaxies in each bin.
}\end{figure}

\begin{figure}[tb]
\plotone{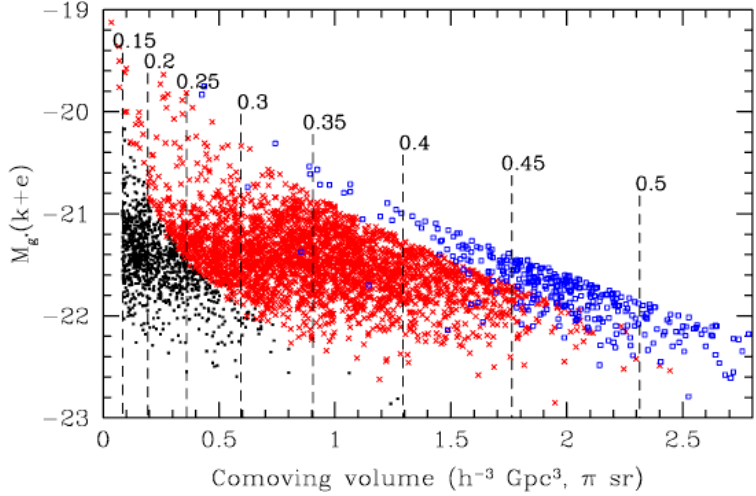} 
\caption{\label{fig:volMg}%
Redshift versus $z=0$ absolute $\gs$ magnitude for 4500 LRGs.  The abscissa
has been remapped to show the enclosed comoving volume for a survey of
$\pi$ steradians as a function
of redshift.  Dashed lines show increments of redshift.
The ordinate has been passively-evolved and $K$-corrected
to $z=0$.  This means that a sample of constant comoving volume will
have an even density of objects.  LRGs from MAIN, Cut I, and Cut II are
shown in black dots, red crosses, and blue squares, respectively.  The
apparent magnitude cuts are clearly visible as the diagonal transitions.
}\end{figure}

\begin{figure}[tb]
\plotone{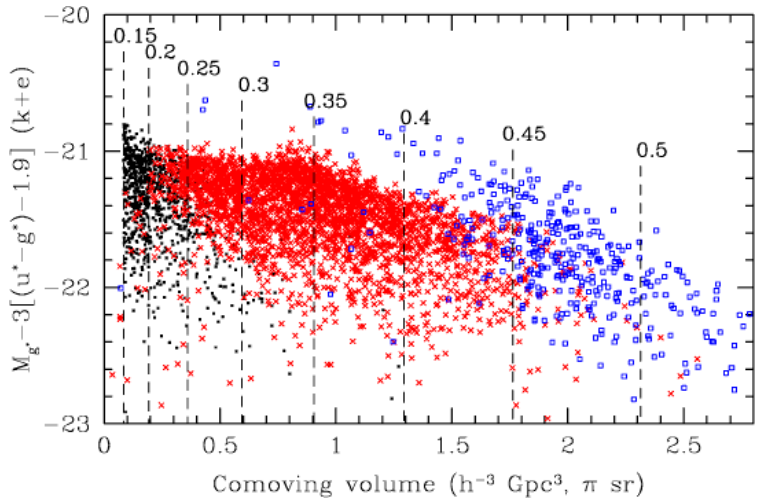} 
\caption{\label{fig:volMgug}%
As Figure \protect\ref{fig:volMg}, but the ordinate has
been replaced by a combination of absolute magnitude and rest-frame
color (both $K+e$ corrected to $z=0$) that lies parallel to the 
LRG boundary line in Figure \protect\ref{fig:Mgug0cutI}.  
This shows that the boundary of the selection is rather sharp and
moves only slightly with redshift for $z<0.4$.
}\end{figure}

\begin{figure}[tb]
\plotone{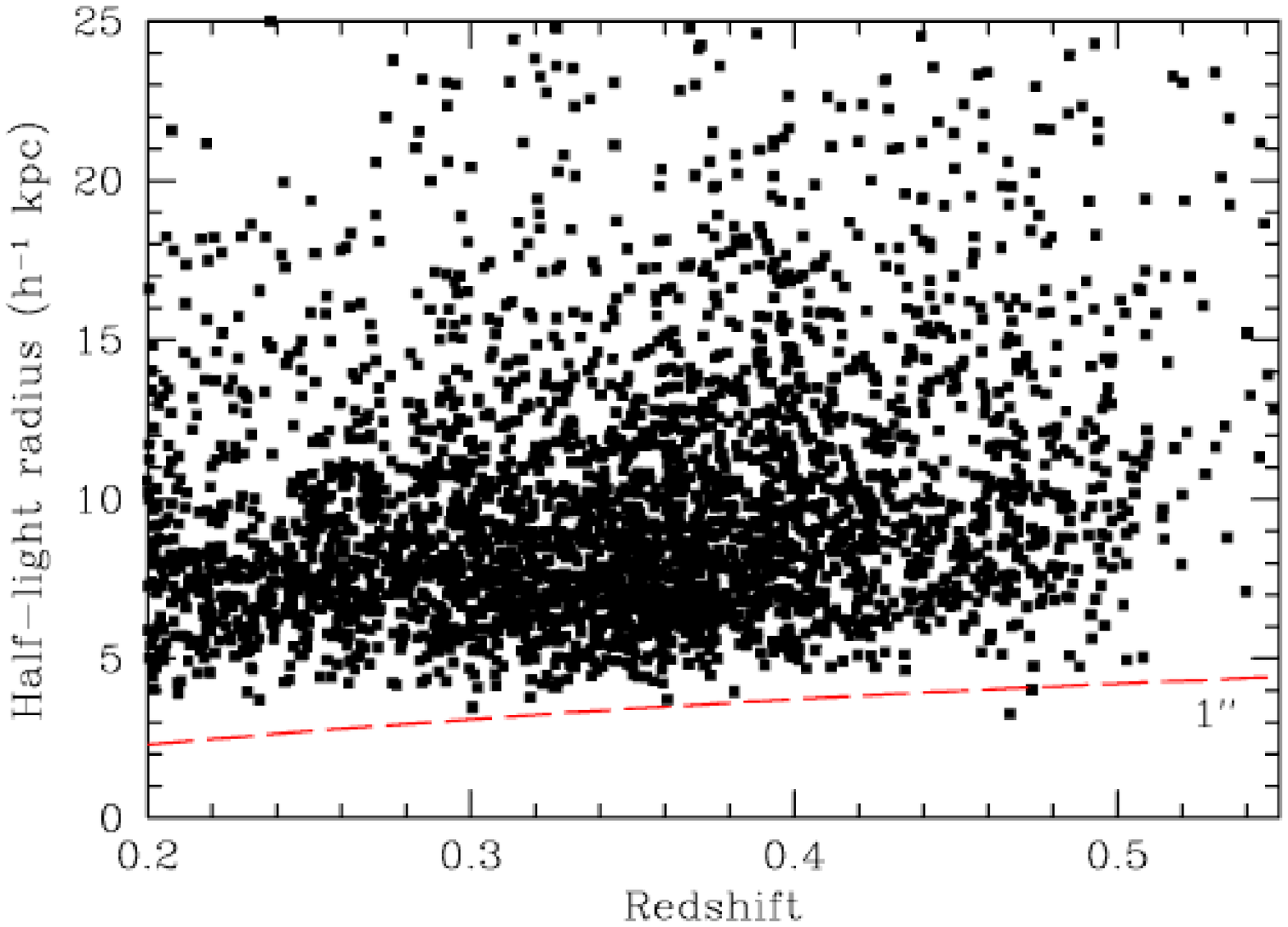} 
\caption{\label{fig:zzsize}%
The effective radius of the best-fit de Vaucouleurs model in the $r$ band
versus redshift for LRGs.
While the model fit is done including the convolution of the point spread
function, we only include objects with $r$ band seeing better than 1$''$.4 just
to minimize any seeing effect.  The dashed line shows the length 
corresponding to 1$''$.
The median radius increases slightly with redshift;  
one possible cause of this might be the increasing luminosity 
threshold at $z>0.4$.  
}\end{figure}

A design goal of the sample is to produce a nearly volume-limited set of
luminous, red galaxies.  This concept cannot be precisely realized
in the context of a merging galaxy population, but we can offer a few
results that are suggestive of a volume-limited sample.
In particular, we study
the comoving density and passively-evolved luminosity thresholds as
functions of redshift and show that these are nearly constant for $z<0.4$.

Figure \ref{fig:zzhist} shows the comoving number density of 
LRGs from 173 plates.  
The sample has a nearly constant density to $z\approx0.4$,
with dropping density beyond.  

Figure \ref{fig:volMg} shows the distribution of redshift and
absolute magnitude.  The latter has been corrected for passive
evolution of an old stellar population.  The redshift axis has
been warped so as to show the enclosed comoving volume; this 
means that a sample of constant comoving density would have a 
constant density of points.  The lines of constant apparent
magnitude are clearly seen as the transition between different types of points.
The sample appears to have a fairly constant luminosity threshold
to $z\approx0.35$, at which point the flux limits of the cuts
impose a floor.
However, the blurred appearance of the luminosity cut
is not caused by measurement errors but rather by the variation
of the cut with rest-frame color, as shown in Figure \ref{fig:Mgug0cutI}.

We can correct for this correlation of luminosity and color by 
measuring the offset of the galaxies from the diagonal boundary
in Figure \ref{fig:Mgug0use}.  This is shown in Figure \ref{fig:volMgug}.
Here we see that the boundary is fairly sharp.  The boundary
appears to be fairly constant for $0.2<z<0.4$.  At larger redshifts,
Cut I has a clear downturn, while Cut II seems to be lower in 
comoving density.  The upturn near $z\approx0.15$
is real and will be discussed in \S\ \ref{sec:main}. 

In addition to approximately constant luminosities, the LRGs also show
constant physical size out to $z\approx0.4$, as shown
in Figure \ref{fig:zzsize}.  The increase in size at $z>0.4$
is expected because those galaxies are more luminous on average.
We have used the effective radius of the best-fit 
seeing-convolved de Vaucouleurs model to define the size.

In summary, the LRG sample appears to have approximately
constant passively-evolved
selection, physical size, and comoving number density 
out to $z\approx0.4$.
From this, we would say that the sample is approximately volume-limited,
but we caution that small changes in the $K+e$ corrections or assumed
cosmology might 
hide modest differences in the selected galaxies between low and high redshift.
Evolution in color is particularly difficult to constrain within the
uncertainties of the modeling.
However, we do find that primary effects of changing the value of $\Omega_m$ is
to shift all of the number densities and $z=0$ absolute magnitudes uniformly 
without introducing a large offset between redshifts of 0.2 and 0.4.

\subsection{Spectral Quality and Repeatability}

\begin{figure}[tb]
\plotone{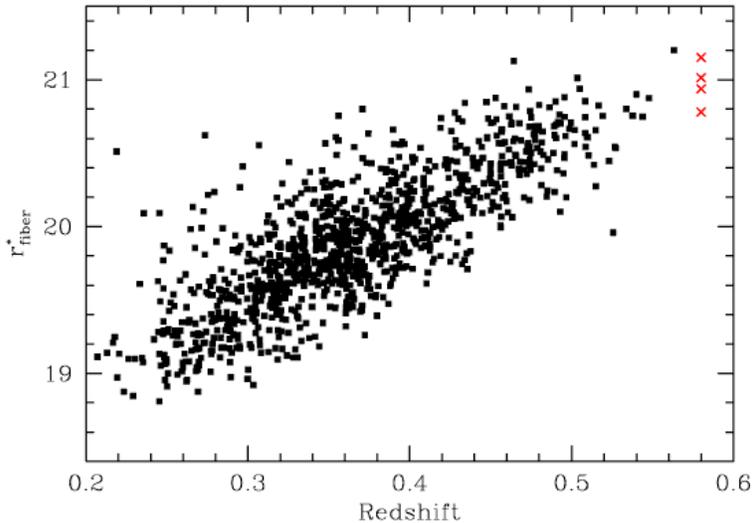} 
\caption{\label{fig:zzrfib}%
Redshift versus 3$''$ aperture (``fiber'') $\rs$ magnitude for LRGs on the
21 verified plates.  The 4 objects for which no redshift was attained
are shown as crosses on the right of the plot.
}\end{figure}

One can use the duplicate observations of the eye-inspected plates 
to quantify the spectral
repeatability.  Here we consider only duplicate observations 
with signal-to-noise ratio exceeding survey minima and only include objects
that were spectroscopically confirmed as extragalactic.  This gave
588 Cut I and 82 Cut II objects with duplicate observations.  The
average signal-to-noise ratio of these duplicates exceeds the minimum 
allowed (by definition)
by 30\%, which is typical of observations in good conditions.

Of the 588 Cut I duplicates, only five differed catastrophically ($>1000$ \kms)
in redshift according to automated software
\citep{Sch01b}.  Three of these five could have their redshifts identified
by eye; the other two suffered from large stretches
of missing spectrum (this is usually due to bad CCD columns).  
Hence, to the eye, one has only two failures,
both because of missing data.
The rms velocity difference of the 583 cases that agreed was 73 \kms
although this drops to 61 \kms\ if one excludes nine outliers with 
velocity differences between 250 \kms\ and 500 \kms.  Note that the uncertainty
in a single measurement would be $\sqrt2$ lower than this if the
two measurements had equal errors; the correction here would be
somewhat less because we are interested in the error of the lower
quality measurement.  One would reasonably say that the redshift
error is around 50 \kms\ rms.  Note that survey performance on the 
MAIN galaxy sample is considerably better than this because the
targets are brighter \citep{Str01}.

Of the 82 Cut II duplicates, one again finds five catastrophic errors 
($>1000$ \kms) from the automated software.  Three were recovered by eye; two were 
too low in signal-to-noise ratio to securely state the redshift.  The 
rms velocity difference of the remaining 77 was 150 \kms, but this
drops to 113 \kms\ if three outliers of more than 350 \kms\ are dropped.
Obviously, errors of several hundred \kms\ are severe and indicate
that the precise redshift is being based on a small number of noisy
absorption lines while the crude redshift is constrained by the
4000\AA\ break.  Again, the above rms differences should be divided
by a factor just shy of $\sqrt2$, so one could claim rms errors in Cut II 
of about 100 \kms.

The LRGs are among the faintest objects targeted by the SDSS,
and we expect that the faintest of the LRGs will have incompleteness
due to inadequate signal-to-noise ratio in the spectra.
Figure \ref{fig:zzrfib} shows the distribution of redshift versus
3$''$ diameter aperture magnitudes (3$''$ is the input diameter of 
the spectroscopic
fiber).  The four objects for which we fail
to get a redshift fall at the faint end of the magnitude distribution.

\section{Using the LRG Sample}
\label{sec:uselrg}

The SDSS \target\ pipeline sets two flags for the LRG sample.  The
GALAXY\_RED flag is set if the object passes either Cut I or Cut II.
The GALAXY\_RED\_II flag is set if the object passes Cut II but not Cut I.
If an object is brighter than the MAIN sample flux cut, then neither
flag is set if the object failed to enter the MAIN sample (for example,
because of the MAIN sample surface brightness cut).  In other
words, LRG target selection never overrules MAIN target selection on
brighter objects.  

As will be described below, the LRG cuts do not preserve the luminosity
threshold at $z<0.2$.  Therefore, the simplest prescription for using the LRG
sample is to select objects with the GALAXY\_RED flag and redshifts
$z>0.2$.  

\subsection{Extracting low-redshift LRGs from the MAIN sample}
\label{sec:main}

The LRG flags are set when a MAIN galaxy passes either of the 
LRG cuts.  This may lead to the impression that such
galaxies are physically similar to the LRGs at higher redshift.
However, this is incorrect because the LRG selection cuts were
not designed to track the color-magnitude locus of luminous,
red galaxies to lower redshifts.  Indeed, such a prescription
would be essentially impossible as $z\rightarrow0$ because
the observed color would not change while the distance modulus
would diverge
and hence the absolute magnitude would become unbounded.  The sense
of the breakdown of the LRG cuts is that they become too permissive;
the linear color-magnitude cut (eq.~[\ref{eq:cutI}]) allows 
underluminous galaxies to enter the sample.

It is of course necessary to include the LRGs from MAIN at
$z\lesssim 0.3$ because the more luminous LRGs are above the MAIN
flux cut, as shown in Figures \ref{fig:zzhist} and \ref{fig:volMg}.
In detail, the imposed color-magnitude cut generates a constant 
passively-evolving luminosity selection down to $z=0.2$.
LRG-selected galaxies with $0.15<z<0.2$ could
be used, but the luminosity cut will be fainter by about 0.1 magnitudes.
By $z=0.1$, the cut has moved by 0.5 magnitudes.  {\it We strongly
advise the reader that the LRG target flags cannot be used 
to select a volume-limited sample at $z<0.15$.}

At redshifts below 0.2, all the galaxies that satisfy the LRG luminosity
threshold are brighter than the MAIN flux cut and therefore are
targeted for spectroscopy regardless of their color.
One is therefore free to select a sample of galaxies based on the
spectroscopic redshift and an appropriate range of (spectroscopically
informed) absolute magnitude and rest-frame color.
Of course, both color and magnitudes must be adjusted for evolution if one 
wants to select the same population across redshift.
We do not have a clean prescription for this at this time; as one can
see by the offset between the low and high redshift loci 
in Figure \ref{fig:Mgug0use}, our $K+e$ corrections do not match
up in rest-frame color, for a variety of reasons explained in 
Appendix \ref{sec:appendix}.  We hope to improve this modeling in 
future work.

\subsection{Caveats and Selection}
\label{sec:caveats}

\begin{figure}[tbp]
\plotone{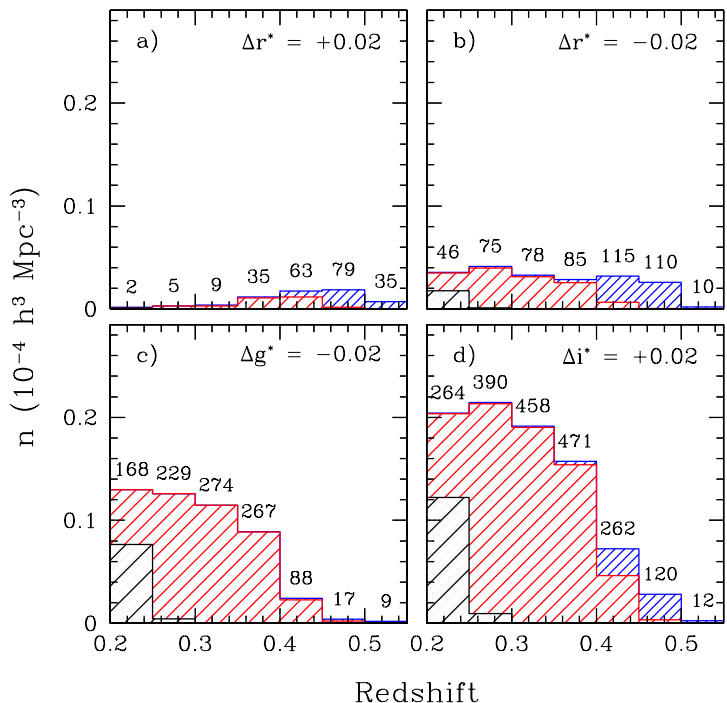} 
\caption{\label{fig:zzhistderiv}%
Redshift distribution of the spectroscopic LRGs 
that are excluded when the zero-point of the magnitude system
is perturbed in the following ways.  
({\it a}) $\rs$ magnitudes fainter by 0.02 mag.
({\it b}) $\rs$ magnitudes brighter by 0.02 mag.
({\it c}) $\gs$ magnitudes brighter by 0.02 mag.
({\it d}) $\is$ magnitudes fainter by 0.02 mag.
The $\rs$ changes affect $\rpet$ magnitudes, $\sbr$ surface
brightness, and model colors;
the $\gs$ and $\is$ changes only enter through the colors.
One may compare these number densities to those in Figure 
\protect\ref{fig:zzhist}, though note the numerical change in 
the vertical axis.  Making $\gs$ fainter or $\is$ brighter
excludes no galaxies.  The effect on the different cuts
are shown with different shading, with MAIN, Cut I, and Cut II
shown left-to-right.
}\end{figure}

The LRG sample selection is extremely sensitive to the calibration
of the $\gs$, $\rs$, and $\is$ photometry.  Figure
\ref{fig:zzhistderiv} shows the redshift distribution of galaxies 
that would be excluded were the photometric zeropoints perturbed
by 0.02 mag.  Making $\is$ fainter by 0.02 mag excludes 20\% of the
galaxies, while making $\gs$ brighter by 0.02 mag excludes 10\%.
Changes in $\rs$ exclude fewer galaxies.  This sensitivity is not
surprising, as one is sampling a very steep portion of the luminosity
function.

Moreover, as apparent in Figure \ref{fig:zzhistderiv},
changes in the photometric zeropoints interact with the
selection cuts to skew the redshift distribution of the sample.
Cut I is primarily affected by the shifting of $\cpar$ and $\rpet$
in equation (\ref{eq:cutI}).  The effect is such that 0.01 mag
in $\cpar$ or 0.03 mag in $\rpet$ adjusts the number density 
by 10\%.  However, to first approximation, this affects all redshifts
equally as one changes the luminosity threshold.

Cut II, on the other hand, is fairly insensitive to changes in $\gs$ 
and hence to $\gs-\rs$.  It is quite sensitive to $\rs$ and $\is$ changes.
The sensitive boundary here is equation (\ref{eq:cutIIcperp}), which
separates the $z>0.4$ region from the low-redshift locus.  Many
Cut II galaxies are near this boundary and near the flux limit.
A blue-ward shift in $\rs-\is$ can leave galaxies outside the boundary, and
if they have $19.2<\rpet<19.5$ then they are too faint for Cut I.
On the other hand, simply moving the $\rs$ zeropoint fainter causes
a noticeable number of Cut II objects to miss the $\rpet<19.5$ flux
cut (Fig.~\ref{fig:zzhistderiv}a).

The combination of the sensitivity to certain offsets in color 
and the measurement errors on these colors will introduce a 
Malmquist-like systematic bias in the measured colors relative to the true 
colors.
For example,
the median quoted error on $\cpar$ is 0.023 mag at $\rs=18$ and 
0.053 mag at $\rs=19$.  
A simple calculation shows that if a population of sources has
a true distribution of a property $x$ proportional to $e^{-\alpha x}$
and that $x$ is measured with a Gaussian error $\sigma$, then 
the apparent number of sources with $x>0$ is inflated by a value
$\exp(\alpha^2 \sigma^2/2)$ and the distribution of true $x$ for
a given observed $x$ is shifted by $-\alpha\sigma^2$.  If $x$
is our color $\cpar$ for Cut I, then $\alpha$ is about 10~mag$^{-1}$,
which means that measurement errors at $\rs=19$, where $\sigma\approx0.05$, 
have inflated the number of targets by 13\% and biased the mean color 
by 0.025 mag. The effects are considerably smaller at $\rs=18$.
We have neglected these biases because they are smaller than our
uncertainties in the $K+e$ corrections, but eventually a detailed 
analysis of the evolution of colors and number densities of LRGs
will need to account for these measurement biases.

It is interesting to note that errors in the amplitude 
of the correction for interstellar extinction have relatively little effect 
on Cut I.  Galaxies become both redder and fainter, and the two effects
partially cancel.  Of course, errors in the slope of the extinction
curve would alter this balance.

\section{Conclusions}
\label{sec:concl}

The luminous red galaxy subsample of the SDSS galaxy spectroscopic survey
will provide a sample of over 100,000
intrinsically luminous early-type galaxies to 
$z\approx0.5$.  The selection is designed to impose a passively-evolving
luminosity cut so as to approach a volume-limited sample.  We have
demonstrated that this holds to $z=0.38$; the sample
becomes flux-limited at higher redshifts because of the signal-to-noise ratio
limits of the spectroscopic data.  The efficiency of the selection at
finding luminous early-type galaxies is very high, about 95\% in
Cut I and 90\% in Cut II.  The success rate for the spectra to yield
a redshift is also very high but may drop for the faintest, highest
redshift galaxies.

The primary science drivers of the LRG sample are to trace clusters of
galaxies out to $z=0.5$ and to provide an enormous volume for the
study of large-scale structure.  It should also provide a large
sample for the study of the evolution of giant ellipticals, provided
that one can account for the effects of the color selection.
As of June 2001, there are 15,000 unique LRGs at $z>0.2$ with survey quality
data in hand, along with over 3000 duplicate observations.
Thus, the SDSS LRG sample currently maps a comoving volume of 
about $1.5\times10^8\hmpcC$ with a relatively uniform set of luminous,
early-type galaxies.  The full sample will cover over $1\hgpcC$.
It is clear from the data in hand that the sample will realize its goal of
providing a powerful extension of the SDSS spectroscopic survey for the
study of structure and galaxy evolution at intermediate redshifts.

\bigskip
The Sloan Digital Sky Survey (SDSS) is a joint project of 
The University of Chicago,
Fermilab, the Institute for Advanced Study, the Japan Participation Group, 
The Johns
Hopkins University, the Max-Planck-Institute for Astronomy (MPIA), the
Max-Planck-Institute for Astrophysics (MPA), New Mexico State University,
Princeton University, the United States Naval Observatory, and the University of
Washington. Apache Point Observatory, site of the SDSS telescopes, is operated by
the Astrophysical Research Consortium (ARC). 

Funding for the project has been provided by the Alfred P. Sloan Foundation, 
the SDSS member institutions, the National Aeronautics and Space Administration, 
the National
Science Foundation, the U.S. Department of Energy, the Japanese Monbukagakusho,
and the Max Planck Society. The SDSS Web site is http://www.sdss.org/. 

We thank Michael Blanton, Doug Finkbeiner, and Ann Zabludoff
for useful conversations, Rob Kennicutt and Rolf Jansen
for supplying their spectrophotometry in electronic form,
and St\'ephane Charlot for supplying his latest population synthesis models.
D.J.E. was supported by NASA through Hubble
Fellowship grant \#HF-01118.01-99A from the Space Telescope Science
Institute, which is operated by the Association of Universities
for Research in Astronomy, Inc, under NASA contract NAS5-26555, as
well as by the Frank and Peggy Taplin Membership at the IAS.

\appendix

\section{Target Selection in the Commissioning Data}
\label{sec:commissioning}

The selection cuts described in section \ref{sec:cuts} 
are current versions and are expected
to be applied throughout the remainder of the SDSS.  However, the commissioning
data, such as those contained in the Early Data Release \citep{Sto01},
were used to refine target selection and hence
applied slightly different cuts as we learned about the 
imaging data and the sample.  
These changes are tracked by the version number of \target.  

Working backwards in time, \target\ versions earlier than v2\_13\_4
(chunks 4 to 9, tiles 73 to 219\footnote{The imaging survey is divided
into disjoint ``chunks'', in which a single instance of \target\ is run
to generate a homogeneous sample within that region of sky.  The resulting
targets are divided onto a number of discrete spectroscopic pointings,
known as ``tiles'', which in turn are realized as ``plates''.  Because
the location of the drilled holes depends on the hour angle of the 
observation, it is possible for multiple plates to be drilled for a given
tile.  See \protect\citet{Bla01b} for more details.})
used a value of 0.3 in the Cut II star-galaxy cut (eq.~[\ref{eq:cutIIsg}]).
Note that this yields a superset of the current selection, so one can 
easily restore the current selection.
\target\ versions earlier than v2\_9 (chunks 4 to 8, tiles 73 to 205) 
used a value of 1.35 in the Cut II
equation (\ref{eq:cutIIgr}) and did not apply the $\rs-\is<1.5$ cut.
\target\ versions earlier than v2\_7 (chunks 4 to 7, tiles 73 to 159) 
applied a Cut II surface brightness
cut (eq.~[\ref{eq:cutIIsbr}]) of 23.3 rather than 24.2; this eliminates
about 15\% of the objects that would now pass Cut II.

Finally, prior to \target\ v2\_7 (chunks 4 to 7, tiles 73 to 159), 
both cuts used the \photo\ variable {\tt objc\_type}
equal to 3 for star-galaxy separation, replacing equations (\ref{eq:cutIsg})
and (\ref{eq:cutIIsg}).  Note that this is the only change to Cut I, but
it has no effect on completeness: since the change, no Cut I object and 
only one Cut II object that turned out to have a non-stellar redshift has 
had {\tt objc\_type} not equal to 3.  For Cut I, only one object out of
$\sim\!3000$ in Chunks 4 to 7 failed equation (\ref{eq:cutIsg}).  Hence,
for Cut I, the change in star-galaxy separation did not affect
the selection of galaxies in any important way.
For Cut II, about
1\% of galaxies in Chunks 4 to 9 fail equation (\ref{eq:cutIIsg}).
The change in the star-galaxy separation cut did somewhat affect
the fraction of stellar interlopers and there are stars that pass the
{\tt objc\_type} cut and fail the $\rpsf-\rmodel$ cut and vice versa.
The variable {\tt objc\_type} was eliminated from the selection to
remove dependences on the $u$ and $z$ bands.

It is important to note that the evolution of the \photo\ pipeline
could also affect the selection function.  In particular, changes in 
the measurement of model magnitudes and the parameters of the 
deblending algorithm may cause subtle differences in the performance
of LRG selection between different chunks.  We have not yet studied
these issues.  The photometric calibration of the commissioning data
was also preliminary; this may affect the selection in the manner
displayed in Figure \ref{fig:zzhistderiv}.

For completeness, we note that the now-obsolete chunk 2
used a completely different set of LRG cuts that we will not document here.
This portion of sky was re-targeted as chunk 8.

\section{$K$-corrections and Passive Evolution}
\label{sec:appendix}

Because the LRG sample spans a wide range of redshifts, the interpretations
of the sample often require the application of $K$-corrections and 
stellar population evolution corrections for comparison of photometry at
different redshifts.  While $K$-corrections alone
could be derived empirically from spectrophotometry, evolutionary corrections
necessarily require models.  In this paper, we use models based
on the PEGASE stellar population model \citep{Fio97} to derive a set of $K+e$
corrections.  

If we only wanted to estimate the passively-evolved absolute magnitude
for an old stellar population, 
then we could rely on a single star formation history, namely
an old single burst.  However, because we also want to investigate
the dispersion in rest-frame colors, we need to specify a range 
of star formation histories.

At the typical redshift of the LRG sample, the observed $g$ and $r$
bands are close to the rest-frame $u$ and $g$ bands, respectively.
Hence, we use $\rpet$ to predict the rest-frame $g^*_{\rm Petro}$ 
magnitude.  For colors, however, we use the observed $\gs-\is$
color to predict the rest-frame $\us-\gs$ color; we pick the wider
baseline so as to avoid concerns about the 4000\AA\ band being in the 
$r$ band at the higher redshift end of the sample.

\begin{figure}[tb]
\plotone{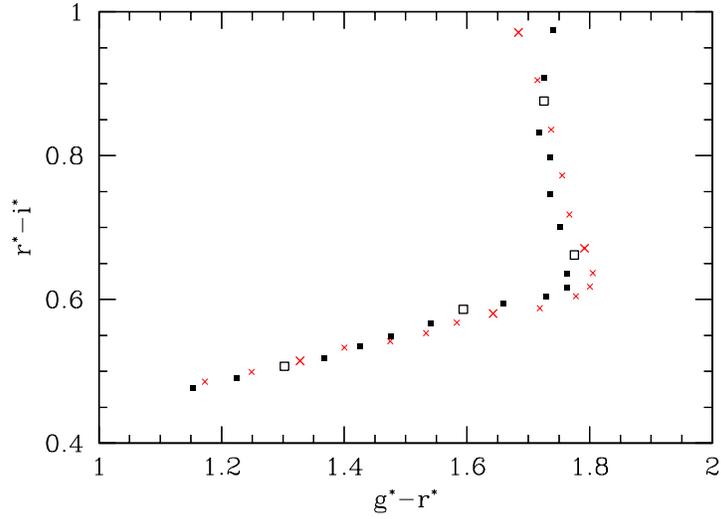} 
\caption{\label{fig:grrizbin}%
Median observed colors of LRGs in a series of redshift bins.  
({\it black}) Each dot indicates the median colors within a slice 
of $\Delta z=0.02$.  Larger dots are slices centered at $z=0.2$, 0.3, 0.4,
and 0.5 (blue to red); smaller dots are centered every 0.02.
({\it red}) Crosses mark the predicted color of an old stellar population
as described in Appendix \protect\ref{sec:appendix}.
}\end{figure}

\begin{figure}[tb]
\plotone{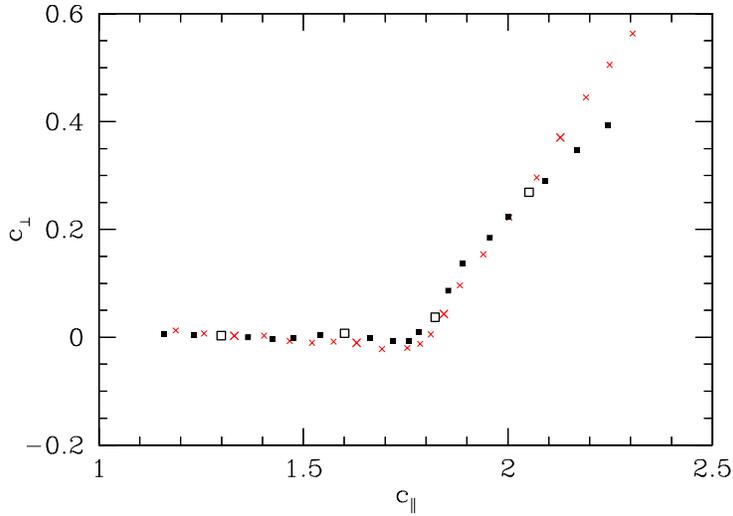} 
\caption{\label{fig:cparcperpzbin}%
As Figure \protect\ref{fig:grrizbin}, but the axes have been
rotated to the $\cpar$--$\cperp$ system.
}\end{figure}

\begin{table}[tb]\footnotesize
\caption{\label{tab:kecorr}}
\begin{center}
{\sc Colors and $K$-corrections for two evolving galaxy models\\}
\begin{tabular}{ccccc\colskip cccc}
\tableskip\hline\hline\tableskip
&\multicolumn{4}{c}{Non-star-forming}&\multicolumn{4}{c}{Star-forming} \\
$z$ & $\Delta\gs$ & $\us-\gs$ & $\gs-\rs$ & $\rs-\is$ & 
	$\Delta\gs$ & $\us-\gs$ & $\gs-\rs$ & $\rs-\is$ \\
\tableskip\hline\tableskip
0.00 & 0.000 & 1.929 & 0.775 & 0.387 &   0.000 & 1.758 & 0.727 & 0.374 \\ 
0.02 & 0.039 & 1.928 & 0.810 & 0.389 &   0.034 & 1.754 & 0.759 & 0.375 \\ 
0.04 & 0.081 & 1.940 & 0.843 & 0.403 &   0.071 & 1.757 & 0.788 & 0.388 \\ 
0.06 & 0.128 & 1.955 & 0.881 & 0.417 &   0.113 & 1.756 & 0.822 & 0.401 \\ 
0.08 & 0.182 & 1.965 & 0.924 & 0.432 &   0.161 & 1.748 & 0.860 & 0.415 \\ 
0.10 & 0.249 & 1.961 & 0.977 & 0.440 &   0.221 & 1.727 & 0.907 & 0.421 \\ 
0.12 & 0.322 & 1.957 & 1.036 & 0.451 &   0.286 & 1.704 & 0.960 & 0.432 \\ 
0.14 & 0.402 & 1.953 & 1.102 & 0.469 &   0.358 & 1.677 & 1.019 & 0.448 \\ 
0.16 & 0.487 & 1.957 & 1.173 & 0.486 &   0.433 & 1.655 & 1.082 & 0.464 \\ 
0.18 & 0.575 & 1.964 & 1.249 & 0.499 &   0.511 & 1.631 & 1.149 & 0.475 \\ 
0.20 & 0.665 & 1.969 & 1.328 & 0.515 &   0.591 & 1.603 & 1.218 & 0.489 \\ 
0.22 & 0.752 & 1.976 & 1.400 & 0.533 &   0.666 & 1.575 & 1.281 & 0.505 \\ 
0.24 & 0.836 & 1.995 & 1.475 & 0.542 &   0.738 & 1.552 & 1.345 & 0.513 \\ 
0.26 & 0.912 & 2.030 & 1.533 & 0.553 &   0.804 & 1.535 & 1.397 & 0.522 \\ 
0.28 & 0.980 & 2.069 & 1.583 & 0.568 &   0.865 & 1.517 & 1.440 & 0.535 \\ 
0.30 & 1.056 & 2.109 & 1.642 & 0.581 &   0.929 & 1.494 & 1.491 & 0.545 \\ 
0.32 & 1.146 & 2.147 & 1.719 & 0.588 &   1.005 & 1.459 & 1.555 & 0.551 \\ 
0.34 & 1.233 & 2.185 & 1.778 & 0.605 &   1.077 & 1.421 & 1.604 & 0.565 \\ 
0.36 & 1.285 & 2.248 & 1.800 & 0.618 &   1.120 & 1.402 & 1.621 & 0.575 \\ 
0.38 & 1.322 & 2.312 & 1.805 & 0.637 &   1.150 & 1.383 & 1.623 & 0.591 \\ 
0.40 & 1.350 & 2.386 & 1.792 & 0.671 &   1.172 & 1.369 & 1.609 & 0.621 \\ 
0.42 & 1.382 & 2.461 & 1.767 & 0.718 &   1.194 & 1.352 & 1.582 & 0.662 \\ 
0.44 & 1.433 & 2.541 & 1.755 & 0.773 &   1.229 & 1.327 & 1.561 & 0.711 \\ 
0.46 & 1.484 & 2.628 & 1.737 & 0.836 &   1.261 & 1.300 & 1.532 & 0.768 \\ 
0.48 & 1.535 & 2.703 & 1.715 & 0.905 &   1.293 & 1.267 & 1.499 & 0.831 \\ 
0.50 & 1.584 & 2.750 & 1.684 & 0.971 &   1.322 & 1.227 & 1.458 & 0.891 \\ 
0.52 & 1.634 & 2.773 & 1.657 & 1.039 &   1.350 & 1.181 & 1.419 & 0.953 \\ 
0.54 & 1.692 & 2.774 & 1.642 & 1.096 &   1.384 & 1.127 & 1.388 & 1.005 \\ 
0.56 & 1.747 & 2.770 & 1.629 & 1.151 &   1.414 & 1.075 & 1.358 & 1.055 \\ 
0.58 & 1.808 & 2.763 & 1.626 & 1.194 &   1.447 & 1.020 & 1.335 & 1.095 \\ 
0.60 & 1.881 & 2.746 & 1.637 & 1.236 &   1.486 & 0.959 & 1.320 & 1.134 \\ 
\tableskip\hline\tableskip
\end{tabular}
\end{center}
NOTES---%
To convert from observed $\rs$ to $z=0$ $M_\gs$, one must subtract the
distance modulus and the value in the $\Delta\gs$ column and add the value
in the $\gs-\rs$ column, all for the redshift in question.  
We linearly interpolate between these two models based on the observed
$\gs-\is$ color of a galaxy at its spectroscopic redshift.
Note that the models have been altered by 0.08 mag blueward in $\gs-\rs$.
This correction is appropriate to the photometric calibrations applied
in \citet{Sto01}, but the models will need to be revisited when new 
calibrations become available.
\end{table}

\begin{figure}[tb]
\plotone{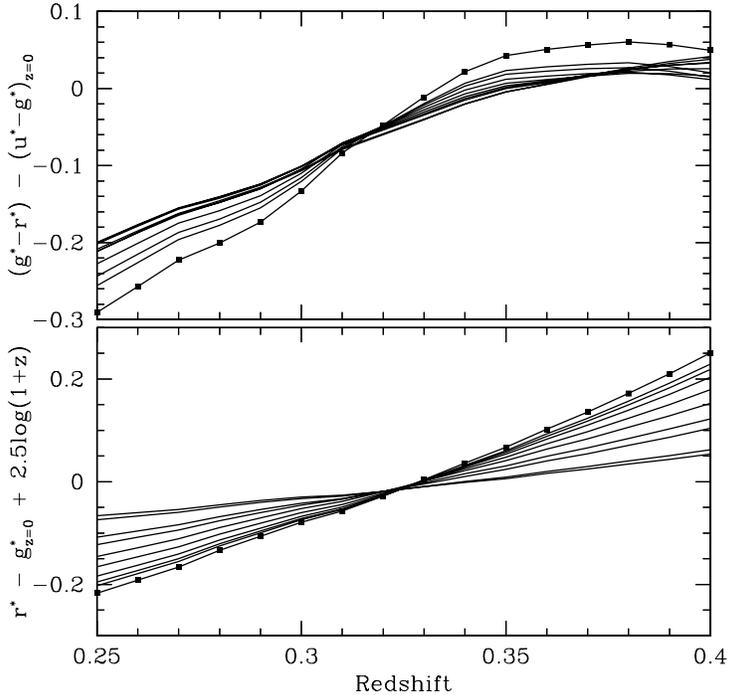} 
\caption{\label{fig:compare_obs_rest}%
({\it top panel}) The comparison of observed $\gs-\rs$ color to 
rest-frame $\us-\rs$ color as a function of redshift for a set
of non-evolving galaxy models.  The models are derived from the PEGASE
code and span a wide range of rest-frame colors; the reddest model
is marked with the large dots.  ({\it bottom panel}) As before, but
for observed $\rs$ magnitude compared to rest-frame $\gs$ magnitude.
We have added a $2.5\log(1+z)$ term so that a perfectly rescaled
filter band would have zero offset in magnitude; this is required
because the definition of the $AB_{95}$ normalization is in energy
per unit frequency \protect\citep{Fuk96}.  
For both the colors and magnitudes, one sees
that the transformation from observed $\gs$ and $\rs$ to rest-frame
$\us$ and $\gs$ is insensitive to the galaxy SED at $z\approx0.32$.
}\end{figure}

\begin{figure}[tb]
\plotone{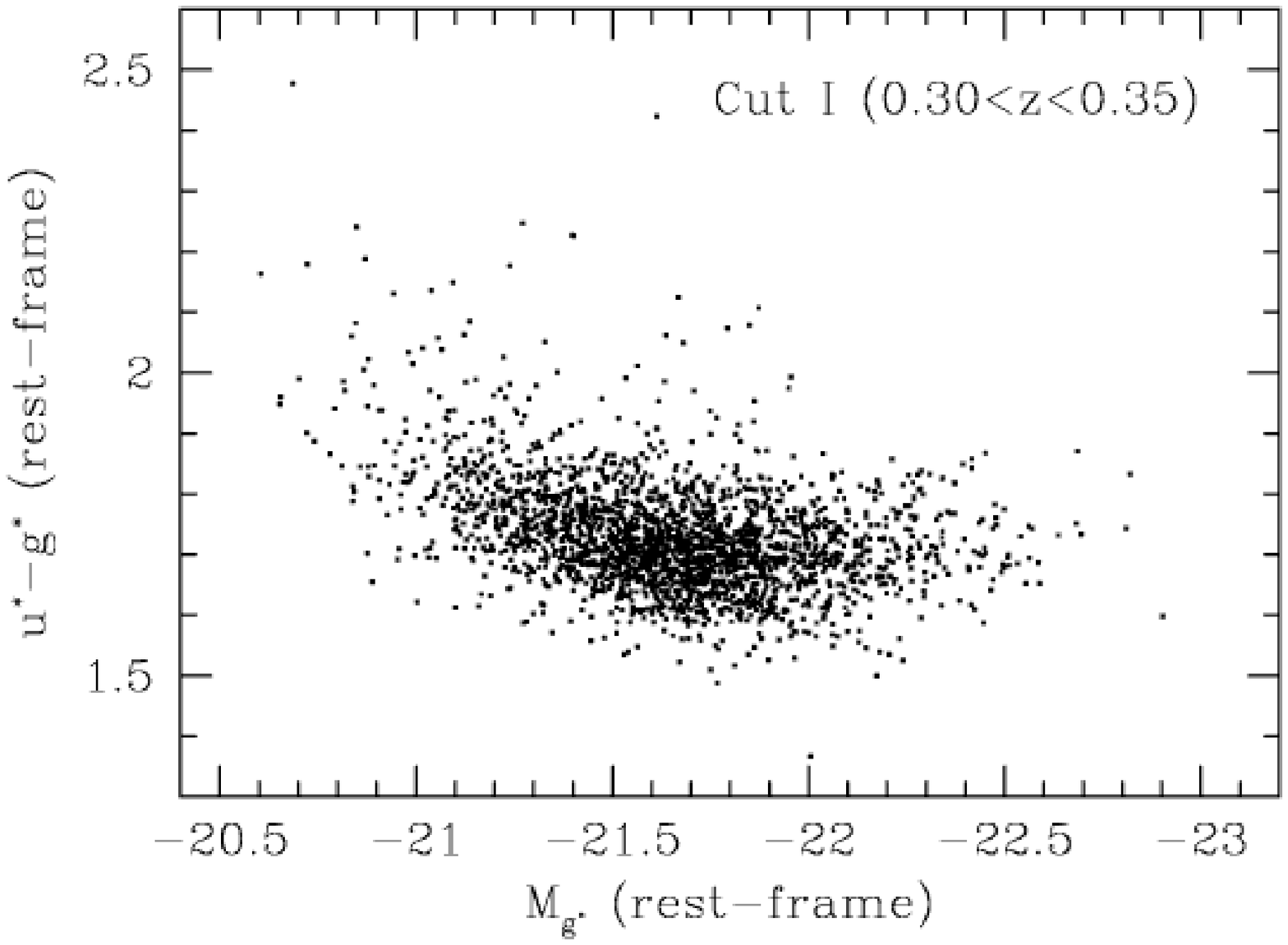} 
\caption{\label{fig:Mg_ug0_z30_35}%
The rest-frame $\us-\gs$ color and absolute $\gs$ magnitude for
LRGs with redshifts between 0.30 and 0.35.  In this range, the
uncertainties of the $K$-corrections are quite small.  No evolution
correction has been applied, and the magnitudes of the models are
not adjusted by the 0.08 mag in $\gs-\rs$ used elsewhere in the
paper.  SDSS calibration uncertainties could still shift the locus.
}\end{figure}

We construct two evolving stellar synthesis models from the PEGASE code.
One is an old, passively evolving burst from $z\approx10$; 
the other is a old population
that quickly forms new stars from the mass loss of existing stars.
Both assume solar metallicity for all stars and a Salpeter initial
mass function.  
Broadband colors are 
calculated using the system response measured for the SDSS \citep{Sto01}.
At any given redshift, the second model is bluer than the first.  We
therefore take the redshift of an LRG and select the linear combination
of the two models that matches the predicted $\gs-\is$ color at that 
redshift to the observed color.
This gives us an effective type, from which we 
quote the $K+e$ corrections to the magnitude and color at $z=0$.

We have encountered a persistent problem 
that all model spectra and
external spectrophotometry of elliptical galaxies 
\citep[e.g.][]{Col80,Ken92,Jan00} that we have tried
predict $\gs-\rs$ colors that are significantly redder (typically about 0.1 mag)
than the SDSS data.
There could be at least three causes: 
the current SDSS photometric calibrations
of one or more bands are not on the $AB_{95}$ system, the assumed 
system response curves are inaccurate in some way, or the external
spectrophotometry has broad-band discrepancies.  However, if the
stellar population synthesis models have such residuals, then the
redshifting of these patterns through the filter bands makes it 
difficult to detect errors in the evolution assumptions.
We do suspect that the SDSS calibration is imperfect but not to the
extent needed to reconcile the galaxy colors.  As a result of the
uncertainties in the calibration and the models, we cannot quote
a set of $K+e$ corrections to the precision needed to compare the
$z=0.4$ galaxies in the LRG sample to the $z=0.1$ galaxies in the
MAIN sample.  

To reconcile the observed color-redshift relation with the two models
used to generate $K+e$ corrections, we subtract 0.08 mag from the $\gs-\rs$
color of the model galaxies.  This puts the bluer model at the lower
envelope of the LRG color data as a function of redshift. 
It is important to note that this affects
the reconstruction of the rest-frame $\us-\gs$ color more than the absolute
$\gs$ magnitude.  The $K+e$ corrections to the 
absolute magnitude in the two models differ by at most 2\%, but
leaving out the color shift causes a noticeable (and non-monotonic) trend 
in rest-frame color versus redshift.  With the shift, the trend is 
much smaller but not absent.

Figure \ref{fig:grrizbin} shows the observed color-redshift relation for LRGs
as compared to the redder model.  Each point marks the median color in $\gs-\rs$
and $\rs-\is$ for a shell in redshift of 0.02 width.
Figure \ref{fig:cparcperpzbin} shows the same figure rotated to the 
$\cpar$-$\cperp$ plane.

While we hope that the modeling of the galaxy colors will improve in
the future, Table \ref{tab:kecorr} presents the colors and magnitudes
of the two models used here (including
the 0.08 mag shift) so that the reader can reproduce the figures in this 
paper as desired.  For reference, the primordial burst model evolves 
0.10 mag in rest-frame $\us-\gs$ and 0.38 mag in rest-frame $\gs$ 
between $z=0$ and $z=0.35$.

We have tried other evolution models for old stellar populations, 
including different parameter sets within PEGASE and models from 
\citet{Bru01}.  We find that the
models generally predict similar amounts of luminosity evolution,
such that we estimate the systematic uncertainty in $M_\gs$ at $z=0$ to be
about 0.1 mag.
However, the models do predict different color-redshift loci, such that 
one could mask or mimic a substantial evolution in color between 
$z\approx 0.4$ and today.  This does not affect the interpretations
of rest-frame color presented in this paper (e.g. the fact that intrinsically
bluer galaxies must be more luminous to pass Cut I), but those who intend
to use the sample to study galaxy evolution will need to confront
this issue.

In Figure \ref{fig:Mgug0use}, we placed the low-redshift MAIN galaxies
on the $M_\gs$ versus rest-frame $\us-\gs$ plane by applying $K+e$
corrections to the observed $\gs$ magnitude and $\us-\gs$ color
(unlike the LRGs, the bright MAIN galaxies have well-measured $\us$
magnitudes).

As a separate exercise, we can minimize our dependence on the modeling
by focusing on a special redshift region. It turns out that the 
spacing of the $\us$, $\gs$, and $\rs$ filters 
is such at $z\approx0.32$ the effective wavelengths of $\gs$ and $\rs$ 
are nearly equal to those of rest-frame $\us$ and $\gs$.  This means that
the observed $\gs-\rs$ color will predict rest-frame $\us-\gs$ with 
little dependence on the SED of the galaxy.
In Figure \ref{fig:compare_obs_rest},
we compare the observed colors to the rest-frame colors (assuming an
${\rm AB}_{95}$ zeropoint) for a variety
of PEGASE models spanning a wide range of star formation histories.
In detail, the $\gs$ filter is fractionally wider than the other 
two bands, which produces a $\sim\!0.05$ mag offset between $\gs-\rs$ and 
rest-frame $\us-\gs$.  One can see that at $z\approx0.32$, the variation
in SEDs makes rather little difference to the conversion to the rest-frame
quantities.

In Figure \ref{fig:Mg_ug0_z30_35}, we use the observed $\gs-\rs$ color 
and $\rpet$ magnitude to calculate the rest-frame $\us-\gs$
color versus absolute $\gs$ magnitude for Cut I LRGs with redshifts
between $0.3$ and $0.35$.  Here, we do not use evolving models
nor do we apply the 0.08 mag shift in $\gs-\rs$ color.  The
Figure therefore shows where LRGs lie in rest-frame color-magnitude 
space at $z\approx0.32$.  Note that to compare to lower redshift
data, one would have to take account of both evolution and any possible
errors in the SDSS photometric zeropoints.

\end{document}